\documentclass[onecolumn,superscriptaddress,amssymb,amsmath,nobibnotes,aps,prd,showpacs,nofootinbib]{revtex4}%
\pdfoutput=1

\usepackage{graphicx}
\usepackage{epsf}
\usepackage{bm}
\usepackage{amsmath}
\usepackage{amsfonts}
\usepackage{amssymb}
\usepackage{epstopdf}
\usepackage{natbib}%
\usepackage{rotating}
\usepackage{pdflscape}
\usepackage{adjustbox}
\usepackage[toc,page]{appendix}

\newcommand{\bea}{\begin{eqnarray}}

\newcommand{\eea}{\end{eqnarray}}
\newcommand{\bma}{\begin{pmatrix}}
\newcommand{\ema}{\end{pmatrix}}
\newcommand{\be}{\begin{equation}}
\newcommand{\ee}{\end{equation}}
\newcommand{\beno}{\begin{equation*}}
\newcommand{\eeno}{\end{equation*}}

\begin{document}

\title{Unification of Dark Matter - Dark Energy in Generalized Galileon Theories}

\author{George Koutsoumbas}
\email{kutsubas@central.ntua.gr}
\affiliation{Department of
Physics, National Technical University of Athens, Zografou Campus
GR 157 73, Athens, Greece}

\author{Konstantinos Ntrekis}
\email{drekosk@central.ntua.gr}
\affiliation{Department of
Physics, National Technical University of Athens, Zografou Campus
GR 157 73, Athens, Greece}

\author{Eleftherios Papantonopoulos}
\email{lpapa@central.ntua.gr}
\affiliation{Department of Physics,
National Technical University of Athens, Zografou Campus GR 157
73, Athens, Greece}

 \author{Emmanuel N. Saridakis}
\email{Emmanuel\_Saridakis@baylor.edu}
\affiliation{Department of
Physics, National Technical University of Athens, Zografou Campus
GR 157 73, Athens, Greece}
\affiliation{Chongqing University of Posts \& Telecommunications, Chongqing, 
400065,China}
\affiliation{CASPER, Physics Department, Baylor University, Waco, TX 
76798-7310, USA}

\begin{abstract}
We present a unified description of the dark matter and the dark energy
sectors, in the framework of shift-symmetric generalized Galileon theories. Considering a 
particular combination of terms in the Horndeski Lagrangian in which we have not  
introduced  a cosmological constant or a matter sector, we obtain an effective unified
cosmic fluid whose equation of state $w_U$ is zero during the whole matter era, namely 
from redshifts $z\sim3000$ up to $z\sim2-3$. Then at smaller redshifts  it starts 
decreasing, passing 
the bound
$w_U=-1/3$, which marks the onset of acceleration, at around $z\sim0.5$. At present times 
  it
acquires the value $w_U=-0.7$. Finally,  it tends toward a de-Sitter
phase in the far future. This behaviour is in excellent agreement with observations.
Additionally, confrontation with Supernovae type Ia data leads to
a very efficient fit. Examining the model at the perturbative level, we show that
 it is free from pathologies such as ghosts and
Laplacian instabilities, at both scalar and tensor sectors, at all times.
\end{abstract}


\maketitle

\section{Introduction}

Recent observations revealed that  the Universe has entered a period of accelerated
expansion \cite{acceleration,ries}. Within the framework of General Relativity (GR), the 
accelerated expansion can be driven by a new energy density component with negative 
pressure, termed Dark Energy (DE) \cite{Copeland:2006wr}, the nature of which is unknown, 
although the cosmological constant might be the
most reasonable candidate. On the other hand, the missing mass in individual galaxies,
as well as their  large scale structure distributions in the whole universe, is attributed
to a new form of matter, termed Cold Dark Matter (CDM), which is assumed to have
negligible pressure. The consideration of these two dark components, along with the usual
standard model of particles, constitutes the concordance model of cosmology, namely the
$\Lambda$CDM paradigm.

Extensions of $\Lambda$CDM concordance model can arise either modifying the gravitational
sector, departing from GR, or maintaining GR but altering the nature of
DE and/or DM. In the first approach, namely of modified gravity (for a review see 
\cite{Joyce:2014kja}), one introduces extra
geometric terms in the action, or extra scalar fields which are non-minimally coupled to
gravity. In the second approach, one considers that the DE sector is dynamic, evolving 
with time, with a possible additional interaction with the DM sector. Such interacting 
dark sectors (for a recent review on the dark matter - dark energy interaction see 
\cite{Wang:2016lxa}) will affect the overall evolution of the Universe and its expansion 
history, the growth of dark matter and  baryon density perturbations, the pattern of 
temperature anisotropies of the Cosmic Microwave Background (CMB) radiation and the 
evolution of the gravitational potential at late times. This approach may also alleviate 
the  cosmic coincidence problem, i.e. the question of why the DE and DM energy densities 
are currently of the same order, although they follow different evolution laws. These
observables are directly linked to the underlying theory of gravity  and, consequently,
the interaction could be constrained by the observational data \cite{Bolotin:2013jpa}.

In a parallel development there has been a large amount of effort  in order to understand 
and
describe the thermal history of the universe in a unified way. Amongst others one could
consider that the  energy density might be regulated by the change in the equation
of state of a background exotic fluid, the Chaplygin gas. The main motivation of
considering  the Chaplygin gas \cite{Bento:2002ps} was to introduce an
equation of state parameter which can mimic a pressureless fluid at the early stages of
the universe evolution, and a quintessence-like sector at late times,
tending asymptotically to a cosmological constant. The generalized Chaplygin gas is
efficient in describing the universe evolution at the background level in agreement with
observations \cite{Bento:2002yx,Bento:2005un}, but it may be plagued by the presence of 
instabilities as well as oscillations
which are not observed in the matter power spectrum \cite{Sandvik:2002jz}. Nevertheless,
one can introduce various
mechanisms in order to bypass such problems, for instance allowing for small entropy
perturbations \cite{Farooq:2010xm,Gorini:2007ta} or for the addition of baryons which can 
improve the behaviour of the matter power spectrum 
\cite{Debnath:2004cd,BouhmadiLopez:2004mp,Setare:2007mp}.

Recently there are many studies of scalar tensor theories \cite{Fujii:2003pa} and one of
them is the Gravity theory resulting from the Horndeski Lagrangian
\cite{Horndeski:1974wa}. Horndeski theories are  technically manageable, since they
lead to second-order field equations, and they prove
consistent  without ghost instabilities \cite{Ostrogradsky:1850fid}. Moreover, a
subclass of these scalar tensor theories of modified gravity shares a classical
Galilean symmetry which has been discovered independently
\cite{Nicolis:2008in,Deffayet:2009wt, Deffayet:2009mn, Deffayet:2011gz}. When one
generalizes Galileon theory abandoning shift symmetry, one reobtains  Horndeski theory.
In Horndeski theory the derivative self-couplings of the scalar field screen the
deviations from GR at high gradient regions (small scales or high
densities) through the Vainshtein mechanism \cite{Vainshtein:1972sx},
thus satisfying solar system and early universe constraints
\cite{Chow:2009fm,DeFelice:2011th,Babichev:2011kq,Saridakis:2016ahq}.

In this work we present a unified description of the dark matter and dark energy sectors
in the framework of shift-symmetric generalized Galileon theory, which is different than 
the standard approach where the Galileon field plays solely the role of dark energy 
\cite{Chow:2009fm,DeFelice:2011th,Babichev:2011kq,
deRham:2011by,Heisenberg:2014kea,Saridakis:2016ahq}, or the approach 
where it plays solely the role of dark matter \cite{Rinaldi:2016oqp}. In particular,
considering a subclass of Horndeski theory, without considering an explicit matter sector
and an explicit cosmological constant, we show that the scalar field gives rise to an
effective cosmic fluid with equation of state parameter that behaves as
pressureless matter at early times, and as dark energy at late times. Additionally, by
suitably choosing the model parameter regions, and without fine tuning, one can obtain the
correct redshift behaviour and values of observables, in agreement with the
observational  data.  Finally, examining the model at the perturbative level we show
that, on top of the background unified solutions, it is free from pathologies such as
ghosts and Laplacian instabilities at all times.

The manuscript is organized as follows. In Section \ref{galmodel} we briefly review the
generalized Galileon cosmology. In Section \ref{DMDEsc} we present the general features
of the dark matter  - dark energy unification in the framework of shift-symmetric
generalized Galileon theories, and in Section \ref{dmdeun} we construct a specific model,
examining its cosmological application in detail. Finally, Section \ref{conclusions} is
devoted to the summary of our results.

\section{Generalized Galileon cosmology}
\label{galmodel}

In this section we briefly review generalized Galileon theory, applying  it to a
cosmological framework and deriving the background equations as well as the
conditions for the absence of instabilities
\cite{DeFelice:2010nf,DeFelice:2011bh}.
  As it is known, in
order to avoid the Ostrogradsky instability \cite{Ostrogradsky:1850fid}
it is required to keep the equations of motion at second order in
derivatives, and thus the most general four-dimensional scalar tensor theories
having second order field equations are described by the action
\begin{equation}
S=\int d^{4}x\sqrt{-g}\,{\cal L}\,,\label{action1}
\end{equation}
 where $g$ is the determinant of the metric $g_{\mu\nu}$, and where the Lagrangian reads
\cite{Horndeski:1974wa}
\begin{equation}
{\cal L}=\sum_{i=2}^{5}{\cal L}_{i}\,,\label{Lagsum}
\end{equation}
 with
\begin{align}
&{\cal L}_{2} = K(\phi,X)~,\label{eachlag2}\\
&{\cal L}_{3} = -G_{3}(\phi,X)\Box\phi~,\\
&{\cal L}_{4} = G_{4}(\phi,X)\,
R+G_{4,X}\,[(\Box\phi)^{2}-(\nabla_{\mu}\nabla_{\nu}\phi)\,(\nabla^{\mu}
\nabla^{\nu}\phi)]\,,\\
&{\cal L}_{5} = G_{5}(\phi,X)\,
G_{\mu\nu}\,(\nabla^{\mu}\nabla^{\nu}\phi)\,\nonumber\\&\ \ \
\ \ \ \ -\frac{1}{6}\,
G_{5,X}\,[(\Box\phi)^{3}-3(\Box\phi)\,(\nabla_{\mu}\nabla_{\nu}\phi)\,
(\nabla^{\mu}\nabla^{\nu}\phi)+2(\nabla^{\mu}\nabla_{\alpha}\phi)\,(\nabla^
{\alpha}\nabla_{\beta}\phi)\,(\nabla^{\beta}\nabla_{\mu}\phi)]\,,\label{
eachlag5}
\end{align}
where for simplicity we have set the gravitational constant to $\kappa\equiv 8\pi G=1$.
The functions $K$ and $G_{i}$ ($i=3,4,5$) depend on the scalar field $\phi$
and its kinetic energy
$X=-\partial^{\mu}\phi\partial_{\mu}\phi/2$, while
$R$ is the Ricci scalar, and $G_{\mu\nu}$ is the Einstein tensor.
$G_{i,X}$ and $G_{i,\phi}$ ($i=3,4,5$) respectively correspond to the
partial
derivatives of $G_{i}$ with respect to $X$ and $\phi$,
namely $G_{i,X}\equiv\partial G_{i}/\partial X$ and
$G_{i,\phi}\equiv\partial
G_{i}/\partial\phi$.

The action (\ref{action1}) was first found by Horndeski in
\cite{Horndeski:1974wa}, but it was independently rederived in the framework of
Galileon theory. We mention here that in the original version of Galileon theory, shift
symmetry plays a crucial role, and hence the two theories do not coincide. Nevertheless,
extending Galileon theory to the so-called generalised Galileon theory, i.e abandoning
the shift symmetry, leads to a complete identification with the Horndeski construction.

Let us now apply the above theory to a cosmological framework. We impose a flat
Friedmann-Robertson-Walker (FRW) background metric of the form
\begin{eqnarray}
ds^{2}=-N^{2}(t)dt^{2}+a^{2}(t)\delta_{ij}dx^{i}dx^{j},
\label{metric}
\end{eqnarray}
where $t$ is the cosmic time,  $x^i$ are the comoving spatial
coordinates, $N(t)$ is the lapse function, and
$a(t)$ is the scale factor. In FRW geometry, $\phi$ becomes a function of $t$ only, and
hence $X(t)=\dot{\phi}^2(t)/2$.

We stress that in standard Galileon cosmology, apart from the
above scalar tensor action (\ref{action1}) one needs to
take into account the matter
content of the universe, described by the Lagrangian
${\cal{L}}_m$, corresponding to a perfect fluid with energy density $\rho_m$
and pressure $p_m$. However, as we mentioned in the Introduction, in the present work we
will not consider an explicit dark matter sector.


 Varying   the action (\ref{action1}) with
respect to   $N(t)$ and $a(t)$ respectively, and setting $N=1$, we obtain
\begin{eqnarray}
   &&
2XK_{,X}-K+6X\dot{\phi}HG_{3,X}-2XG_{3,\phi}-6H^{2}G_{4}+24H^{2}X(G_{4,X}
+XG_{4,XX})\nonumber \\
   &&-12HX\dot{\phi}\, G_{4,\phi X}-6H\dot{\phi}\,
G_{4,\phi}
+2H^{3}X\dot{\phi}\left(5G_{5,X}+2XG_{5,XX}\right)\nonumber \\
   &&-6H^{2}X\left(3G_{5,\phi}
+2XG_{5,\phi X}\right)=0
\,,\label{be1}
\end{eqnarray}
\begin{eqnarray}
   && K-2X(G_{3,\phi}+\ddot{\phi}\,
G_{3,X})+2(3H^{2}+2\dot{H})G_{4}-12H^{2}XG_{4,X}-4H\dot{X}G_{4,X}-8\dot{H}
XG_{4,X}\nonumber \\
  &&-8HX\dot{X}G_{4,XX}
+2(\ddot{\phi}+2H\dot{\phi})G_{4,\phi}+4XG_{4,\phi\phi}+4X(\ddot{\phi}
-2H\dot{\phi})G_{4,\phi
X}\nonumber \\
&&-2X(2H^{3}\dot{\phi}+2H\dot{H}\dot{\phi}+3H^{2}\ddot{\phi})G_{5,X}-4H^{2}
X^{2}\ddot{\phi}\, G_{5,XX} +4HX(\dot{X}-HX)G_{5,\phi
X}\nonumber \\
   &&+2[2(\dot{H}X+H\dot{X})+3H^{2}X]G_{5,\phi}+4HX\dot{\phi}\,
G_{5,\phi\phi}=0
\,,\label{be2}
\end{eqnarray}
 where dots denote derivatives with respect to $t$, and we also
defined the Hubble parameter $H\equiv\dot{a}/a$. Variation of
(\ref{action1}) with respect to $\phi(t)$ provides its evolution equation
\begin{equation}
\frac{1}{a^{3}}\frac{d}{dt}\left(a^{3}J\right)=P_{\phi}\,,\label{fieldeq}
\end{equation}
 with
\begin{eqnarray}
J & \equiv & \dot{\phi}K_{,X}+6HXG_{3,X}-2\dot{\phi}\,
G_{3,\phi}+6H^{2}\dot{\phi}(G_{4,X}+2XG_{4,XX})-12HXG_{4,\phi X}\nonumber
\\
 &  & +2H^{3}X(3G_{5,X}+2XG_{5,XX})-6H^{2}\dot{\phi}(G_{5,\phi}+XG_{5,\phi
X})\,,\\
P_{\phi} & \equiv & K_{,\phi}-2X\left(G_{3,\phi\phi}+\ddot{\phi}\,
G_{3,\phi X}\right)+6(2H^{2}+\dot{H})G_{4,\phi}+6H(\dot{X}+2HX)G_{4,\phi
X}\nonumber \\
 &  & -6H^{2}XG_{5,\phi\phi}+2H^{3}X\dot{\phi}\, G_{5,\phi X}\,.
 \label{Pphidef}
\end{eqnarray}

We close this section by mentioning that in order for the above scenario
to be free of ghosts and Laplacian instabilities, and thus cosmologically
viable, two conditions related to scalar perturbations must be satisfied
\cite{DeFelice:2010pv,DeFelice:2011bh,Appleby:2011aa}. In particular, for the scalar 
perturbations these
read as \cite{DeFelice:2011bh}
\begin{equation}
c_{S}^{2}\equiv\frac{3(2w_{1}^{2}w_{2}H-w_{2}^{2}w_{4}+4w_{1}w_{2}\dot{w}_{1}-2w_{1}^{2}
\dot{w}_{2})
}{w_{1}(4w_{1}w_{3}+9w_{2}^{2})}
\geq0
\label{cscon}~,
\end{equation}
for the avoidance of Laplacian instabilities
associated with the scalar field propagation speed,
 and
  \begin{equation}
Q_{S}\equiv\frac{w_{1}(4w_{1}w_{3}+9w_{2}^{2})}{3w_{2}^{2}}>0\,,
\label{Qscon}
\end{equation}
for the absence of ghosts,
 where
\begin{eqnarray}
&&w_{1}  \equiv  2\,(G_{{4}}-2\,
XG_{{4,X}})-2X\,(G_{{5,X}}{\dot{\phi}}H-G_{{5,\phi}})
\,,
\label{w1def}\\
&&w_{2}  \equiv  -2\, G_{{3,X}}X\dot{\phi}+4\, G_{{4}}H-16\,{X}^{2}G_{{4,{
XX}}}H+4(\dot{\phi}G_{
{4,\phi X}}-4H\, G_{{4,X}})X+2\, G_{{4,\phi}}\dot{\phi}
\nonumber
\\
 &  & \ \ \ \ \ \ \ \,
 +8\,{X}^{2}HG_{{5,\phi X}}+2H\, X\,(6G_{{5,\phi}}-5\,
G_{{5,X}}\dot{\phi}{H})-4G_{{5,{
XX}}}{\dot{\phi}}X^{2}{H}^{2}\,,\\
&&
w_{3}  \equiv  3\, X(K_{,{X}}+2\, XK_{,{  XX}})+6X(3X\dot{\phi}HG_{{3,{
XX}}}-G_{{3,\phi X}}
X-G_{{3,\phi}}+6\, H\dot{\phi}G_{{3,X}})
\nonumber \\
 &  & \ \ \ \ \ \ \ \,
 +18\, H(4\, H{X}^{3}G_{{4,{  XXX}}}-HG_{{4}}-5\, X\dot{\phi}G_{{4,\phi
X}}-G_{{4,\phi}}\dot{
\phi}+7\, HG_{{4,X}}X+16\, H{X}^{2}G_{{4,{  XX}}}-2\,{X}^{2}\dot{\phi}G_{{4,\phi{
XX}}})\nonumber \\
 &  & \ \ \ \ \ \ \ \,    +6{H}^{2}X(2\, H\dot{\phi}G_{{5,{
XXX}}}{X}^{2}-6\,{X}^{2}G_{{5,\phi{  XX}}}+13XH\dot{\phi}G_{{5,{  XX}}}-27G_{{5,\phi
X}}X+15\, H\dot{\phi}G_{{5,X}}-18G_{{5,\phi}})\,,\\
&&w_{4}  \equiv  2G_{4}-2XG_{5,\phi}-2XG_{5,X}\ddot{\phi}~.
\end{eqnarray}
Note that although a negative sound speed square should be obviously avoided, a 
sound speed square larger than one, namely superluminality, does not necessarily imply 
pathologies or acausality around any cosmological 
background \cite{Babichev:2007dw,Deffayet:2010qz}. Instead,
a superluminal propagation around just one possible solution would
imply that the theory cannot be UltraViolet-completed by a local, Lorentz-invariant 
Quantum Field Theory or a weekly coupled string theory \cite{Adams:2006sv}. Hence to 
check the subluminality around one interesting solution (especially around a cosmological 
one) it is not sufficient to make any
definite conclusions about a possible UltraViolet-completion. In particular, in
context of scenarios similar to the one investigated in the present work, 
it was showed in \cite{Easson:2013bda} that if even all 
cosmological solutions are subluminal without external matter, one can still acquire 
superluminal solutions in the presence of normal 
external matter like radiation.

Additionally, for the tensor perturbations the conditions for avoidance of 
  ghost and 
Laplacian instabilities are respectively written as  \cite{DeFelice:2011bh} 
\begin{align}
\label{tens1}
Q_T &= \frac{w_1}{4} >0\\ 
c_T^2 &= \frac{w_4}{w_1}\geq0.
\label{tens2}
\end{align}
Similarly to the scalar perturbations, we mention that although a negative sound 
speed square for the tensor perturbations should be obviously avoided, a 
sound speed square larger than one does not necessarily imply 
pathologies or acausality.

In the next section, using the Einstein equations (\ref{be1}) and (\ref{be2}) and the 
field equation for the scalar field (\ref{fieldeq}), we will  obtain a unified model of
dark energy and dark matter, respecting the above restrictions.

\section{Dark matter - dark energy unification}
\label{DMDEsc}

The main motivation of the present work is to use the scalar degree of freedom of the
 generalized Galileon theory to describe not only the dark energy sector (as it is
the usual approach in Galileon/Horndeski considerations), but to describe both
dark matter and dark energy sectors in a unified way. In particular, without considering
an explicit matter sector, we will use a  combination of terms of the
above theory to construct a specific model, which
behaves as the standard GR theory plus an extra degree of freedom that depends on the 
scalar field. 
This can be expressed as an effective unified fluid that behaves as dark matter at early
and intermediate times, and as dark energy at late times.

 To  obtain an Einstein-frame description,  we need to consider
$G_4(\phi,X) = \frac{1}{2}$ in the action (\ref{action1}), which leads to the appearance 
of
the standard GR term, namely the Ricci scalar. The remaining
separate Lagrangians, i.e $\mathcal{L}_2,\mathcal{L}_3,\mathcal{L}_5$
will be interpreted as parts of the Lagrangian of the unification fluid $\mathcal{L}_U =
\mathcal{L}
_2 + \mathcal{L}_3 + \mathcal{L}_5$. Therefore, in an FRW geometry the Einstein equations
(\ref{be1}) and (\ref{be2})
 reduce to
\begin{align}
3H^{2} &= \rho_U \label{frrho}~, \\
-(3H^{2}+2\dot{H}) &= p_U~,
\label{frpre}
\end{align}
where
\begin{align}
\rho_U &= 2XK_{,X}-K+6X\dot{\phi}HG_{3,X}-2XG_{3,\phi}
+2H^{3}X\dot{\phi}\left(5G_{5,X}+2XG_{5,XX}\right)
   -6H^{2}X\left(3G_{5,\phi}
+2XG_{5,\phi X}\right)\label{rhoUf}~, \\
p_U &= K-2X(G_{3,\phi}+\ddot{\phi}\,
G_{3,X})-2X(2H^{3}\dot{\phi}+2H\dot{H}\dot{\phi}+3H^{2}\ddot{\phi})G_{5,X}-4H^{2}
X^{2}\ddot{\phi}\, G_{5,XX} +4HX(\dot{X}-HX)G_{5,\phi
X}\nonumber \\
&\quad +2[2(\dot{H}X+H\dot{X})+3H^{2}X]G_{5,\phi}+4HX\dot{\phi}\
G_{5,\phi\phi}~. \label{pUf}
\end{align}
The above quantities can be used to define the total equation-of-state parameter of
the Universe $w_U$, that is of the fluid which incorporates both
dark matter and  dark energy sector in a unified way, namely
\be
w_U \equiv \frac{p_U}{\rho_U}~.
\label{totalEoS}
\ee

The goal is to suitably choose the involved functions
$K(\phi,X)$, $G_3(\phi,X)$ and $G_5(\phi,X)$ in order to obtain a behaviour of $w_U$ in
agreement with the observed one. In particular, as it is well known, in standard
cosmology the total equation of state parameter of
the Universe remains very close to zero during the matter dominated era, namely from
redshifts $z\sim3000$ up to $z\sim2-3$ \cite{Ade:2015xua}. Then it starts decreasing,
passing the bound $-1/3$, which marks the onset of acceleration, at around 
$z\sim0.4-0.6$, 
and finally resulting to a value around $-0.7$ at present times (i.e at $z=0$)
\cite{Ade:2015xua}. In standard cosmology, for instance in $\Lambda$CDM paradigm, the 
above behaviour is obtained by the usual consideration of a pressureless dark matter 
fluid 
with equation of state $w_m\approx0$ and
a cosmological constant with equation of state $w_\Lambda=-1$, with corresponding density 
parameters $\Omega_m$ and $\Omega_\Lambda$ respectively. Hence, the total equation of 
state is given by $w_t=\Omega_m w_m+\Omega_\Lambda w_\Lambda\approx \Omega_\Lambda 
w_\Lambda$, and
thus, since $\Omega_\Lambda$ almost vanishes during the matter era, while it starts
dominating only after  $z\sim2-3$, one acquires the above behaviour for the total 
equation 
of state of the Universe.

A crucial consideration for our construction is the assumption of shift symmetry. Under 
shift symmetry, in the equations \eqref{rhoUf}, \eqref{pUf} and
\eqref{fieldeq} only the derivatives  $X(t)$ and
$\dot{X}(t)$ of the scalar field appear, but not the scalar field $\phi$ itself.
This allows to eliminate completely the derivatives of the scalar field between $\rho_U$
and $p_U$, resulting in an expression $p_U(\rho_U)$. Fluids with this kind of equation of 
state, for instance the Chaplygin gas and its extensions \cite{Bento:2002ps}, have been 
shown to be able to induce a unified description of the dark matter and dark energy 
sectors. However, in these models the relation $p_U(\rho_U)$ is arbitrarily set by hand  
while in our model we will show that this relation is obtained from the
theory, and all the functions that appear in this relation are coupling functions  of the 
Horndeski Lagrangian.

The functions $K(\phi,X)$, $G_3(\phi,X)$ and $G_5(\phi,X)$ that appear in the Horndeski 
Lagrangian 
should have the following properties in order to  lead the induced
$w_U$ to have the aforementioned behaviour. Concerning $K(\phi,X)$ shift symmetry requires
to be a function of $X$ only, namely $K(X)$, which includes the canonical kinetic term
$\sim X$. Concerning $G_3(\phi,X)$ the obvious shift-symmetric choice would be
$G_3(\phi,X) = G_3(X)$, however even a term $G_3(\phi,X) \sim \phi$ (which corresponds to 
the 
simplest non-trivial Galileon term $\phi\square\phi$ in the action) leads also to
shift-symmetric equations through integration by parts.

Concerning now
$G_5(\phi,X)$, a term $\sim \phi$ through integration by parts gives rise to the usual
non-minimal derivative coupling term in the action, namely $G_{\mu\nu}\nabla^\mu \phi
\nabla^\nu\phi~$  \cite{Amendola:1993uh}, which is a subclass of
Galileon/Horndeski theory. In particular, the derivative coupling of the scalar field to 
Einstein tensor introduces a new scale in the theory which on short distances allows
to find  black hole solutions \cite{Kolyvaris:2011fk,Rinaldi:2012vy,Kolyvaris:2013zfa},
while if one considers the gravitational collapse of a scalar field coupled to the 
Einstein tensor then a black hole is formed \cite{Koutsoumbas:2015ekk}.
On large distances the presence of the derivative coupling acts as a friction term in 
the inflationary and late-time period of the cosmological
evolution, and hence it was independently introduced and studied with many interesting
cosmological applications 
\cite{Sushkov:2009hk,Gao:2010vr,Granda:2009fh,Saridakis:2010mf,Germani:2010gm,
Dent:2013awa, Dalianis:2016wpu}. Moreover, it was found that at the end of
inflation in the preheating period, there is a suppression of heavy particle production  
as the derivative coupling is increased. This has been  attributed to the fast decrease 
of 
kinetic energy of the scalar field due to its  wild oscillations 
\cite{Koutsoumbas:2013boa}. This change of the kinetic energy of the scalar field coupled 
to Einstein tensor allowed to holographically simulate the effects of a high 
concentration 
of impurities in a material \cite{Kuang:2016edj}.

In this work  we will use this property of the scalar field coupled to Einstein tensor in 
order to restrain the kinetic energy of the scalar field to almost constant values for a 
long time intervals and induce an almost zero $w_U$ during the redshift regime of the
matter era. Then, when the non-minimal derivative coupling weakens, the Universe
expansion makes the field's kinetic energy decrease, and hence $w_U$ departs
towards negative values, signaling the onset of acceleration as required.

We close this section by making a crucial comment concerning the absence of pathologies,
such as ghosts and Laplacian instabilities. It is well known that in principle many
gravitational modifications can have a consistent behaviour at the background level, but
present various pathologies (ghosts, instabilities, super-luminalities etc) at the
perturbative level (as it was the case for example  of basic Ho\v{r}ava-Lifshitz gravity 
\cite{Bogdanos:2009uj} and of basic nonlinear massive gravity \cite{DeFelice:2012mx}). 
Hence, in general, such modified theories 
are free from  pathologies only under specific conditions, which in principle depend on 
both the underlying geometry but also on the given background solution. Thus, after 
obtaining a particular solution, it is crucial to examine the validity of the 
pathologies-absence conditions on-shell of this solution.  In  the present work, after 
investigating solutions that induce a background evolution with the desired features, we 
will verify whether the pathologies-absence conditions (\ref{cscon}) and (\ref{Qscon}) 
are 
indeed satisfied.

In the next Section we present a simple model satisfying all the above requirements,
and we investigate in detail its cosmological implications.

\section{A specific model}
\label{dmdeun}

We will present a specific model along the lines described in the previous Section,
namely construct a specific subclass of shift-symmetric Galileon theories which presents a
unified description of the dark-matter and dark-energy sectors. In particular, we will
consider the canonical kinetic term along with two non-trivial Galileon terms and the
usual non-minimal derivative coupling. Hence, we consider the
Lagrangian
given in (\ref{Lagsum}), with the function choices
 $K(\phi,X) = \frac{X}{2} -\frac{ \eta}{2} X^{1/2}$ ,  $G_3(\phi,X) = 
\frac{\lambda_3}{2}X^{-1/2}$ 
, 
$G_4(\phi,X) =
\frac{1}{
2}$ ,
$G_5(\phi,X) = -\frac{\lambda_5}{2}\phi$, and for convenience we use units where
$8\pi G = c = \hbar = 1$. The action (\ref{action1}) then becomes:
\be
\label{actionfin}
S= \int d^4 x \sqrt{-g}\left[\frac{R}{2} +\frac{1}{2} \left(X -
 \eta X^{1/2}\right) -
\frac{\lambda_3 X^{-1/2}}{2}\square\phi
+ \frac{\lambda_5}{2} G_{\mu\nu}\nabla^\mu\phi \nabla^\nu\phi\right]~.
\ee
Without loss of generality we choose $\eta$ such that its value is 
$\eta=1$. We stress here that the above specific forms of  $K(\phi,X)$ and  
$G_3(\phi,X)$, which contain $X^{\pm1/2}$, are not
very important for obtaining the desired phenomenological cosmological evolution at the 
background level, however they play a crucial role in ensuring the validity of the two
pathologies-absence conditions during the whole evolution, as we will show later on. 
In particular, models with $K \varpropto \sqrt{X}$  correspond to  
$Cuscuton$ scenarios \cite{Afshordi:2006ad}, while models   without $G_5$ and with 
$G_4$ as in usual general relativity correspond the so-called $Kinetic 
\;Gravity \;Braiding$ \cite{Deffayet:2010qz}, and have  been studied independently in the 
literature. These models  exhibit superluminal sound 
speed and thus their addition in the scenario of the present work ensures that the sound 
speed square will increase and avoid negative values. Finally, as 
we mentioned earlier, we do not include a term proportional to $\phi\square\phi$ since 
through integration by parts such a term will give the same contribution to the equations 
of motion as the term proportional to $X$.

In this case, the two Friedmann equations (\ref{be1}) and (\ref{be2}) become respectively
\be  \label{frt}
3H^2   = \left(\frac{1}{2} + 9\lambda_5 H^2\right) X - \frac{3\lambda_3 H}{\sqrt{2}}~,
\ee
and
\be  \label{frt2}
-(3H^{2}+2\dot{H})=
-\frac{1}{2} \sqrt{X} + X \left(\frac{1}{2} - 3\lambda_5 H^2 - 2\lambda_5 \dot{H}\right) 
-2\lambda_
5 H
\dot{X} + \frac{\lambda_3 \dot{X}}{2\sqrt{2}X},
\ee
with $H(t) = \frac{\dot{a}(t)}{a(t)}$ and $X(t) =
\frac{\dot{\phi}^2(t)}{2}$,
 which can be brought to the form of equations (\ref{frrho}) and (\ref{frpre}) with
\begin{eqnarray}
\label{dent}
&&\rho_U\equiv
 \left(\frac{1}{2} + 9\lambda_5 H^2\right) X - \frac{3\lambda_3 H}{\sqrt{2}}~,\\
\label{pret}
&& p_U\equiv
-\frac{1}{2} \sqrt{X} + X \left(\frac{1}{2} - 3\lambda_5 H^2 - 2\lambda_5 \dot{H}\right) 
-2\lambda_
5 H
\dot{X} + \frac{\lambda_3 \dot{X}}{2\sqrt{2}X}~.
 \end{eqnarray}
 Hence, the total equation of state parameter of the Universe \eqref{totalEoS} reads
 \begin{eqnarray}\label{EoS}
w_U\equiv\frac{p_U}{\rho_U}=
\frac{-\frac{1}{2} \sqrt{X} + X \left(\frac{1}{2} - 3\lambda_5 H^2 - 2\lambda_5 
\dot{H}\right)
-2\lambda_5 H \dot{
X} + \frac{\lambda_3 \dot{X}}{2\sqrt{2}X}}{ (\frac{1}{2} + 9\lambda_5 H^2) X - 
\frac{3\lambda_3
H}{\sqrt{2}
}}~.
\end{eqnarray}
In addition, the Klein-Gordon equation  (\ref{fieldeq})  becomes
\begin{align}
-6\sqrt{2}\frac{1}{2} H X^{3/2} &+ 12\sqrt{2}H X^2 \left(\frac{1}{2} + 3\lambda_5 + 2 
\lambda_5
\dot{H}\right) +
3\lambda_3 H \dot{X} + 2 X \left[-3\lambda_3 \dot{H} + \frac{1}{2} \sqrt{2} \dot{X} + H^2
\left(-9\lambda_
3 + 3\sqrt{2}\lambda_5 \dot{X}\right) \right] = 0~,
\label{kgt}
\end{align}
which, using equations \eqref{dent} and \eqref{pret}, can be rewritten in the standard
conservation form, namely
\begin{eqnarray}
\dot{\rho}_U+3H(\rho_U+p_U)  =  0~.
\label{rhoreqde}
\end{eqnarray}
Lastly, we mention that although the function $G_3(\phi,X) = 
\frac{\lambda_3}{2}X^{-1/2}$   
diverges at $X=0$, the equations are well-behaved at this value. Nevertheless, the 
Minkowski vacuum, where $\dot{\phi}=0$, is not a solution of the equations, unless 
$\lambda_3=0$, which is a non-trivial property of the construction. As we mentioned 
earlier, the presence of $\lambda_3 X^{-1/2}$ is not
necessary for obtaining the desired phenomenological evolution at the
background level, but it plays a role in the satisfaction of the 
pathologies-absence conditions. One might search for other $G_3(\phi,X)$ forms, that 
still improve the satisfaction of the pathologies-absence conditions, but they lead to 
the acceptance of the solution of the Minkowski vacuum too.

As we have already stated, the fact that the considered action
(\ref{actionfin}) exhibits the shift symmetry ensures that the scalar field $\phi$ does
not appear in the equations of motion but only its derivatives (i.e $X(t)$ and
$\dot{X}(t)$) do so. This allows to use equations \eqref{frt}, \eqref{kgt} and 
\eqref{dent} in
order to
easily eliminate  these derivatives from equation \eqref{pret}, resulting to an 
expression 
of $p_U$
in terms of $\rho_U$, namely
\begin{small}
\begin{align}
p_U(\rho_U) &= \Big\{[3 \lambda_5 f(\rho_U)-2] 
\left\{-3 \lambda_3^2 \{4 + \lambda_5 
f(\rho_U) [ 9 \lambda_5 f(\rho_U)-28]\} [\sqrt{3}  \lambda_3 +
       g(\rho_U)]
       \right.\nonumber
       \\ &\;\;\;\; \left.+
    2 \frac{1}{2} f(\rho_U) [
       3 \lambda_5 f(\rho_U)-2]
       \left\{\sqrt{3} \lambda_3 \lambda_5 f(\rho_U) [
          3 \lambda_5 f(\rho_U)-14] -
       2 \{2 + 3 \lambda_5 f(\rho_U)[ \lambda_5 f(\rho_U)-1]\} 
g(\rho_U)\right\}\right\}\Big\}^{-
1}\nonumber \\
&\cdot
\Big\{
-36 \sqrt{3}\lambda_3^5 -
  18  \lambda_3^4 \{2 g(\rho_U) + \lambda_5 f(\rho_U) [ \sqrt{3}  \lambda_3 + 
g(\rho_U)]\}  -
  2 \frac{1}{2}^2 f^{3/2}(\rho_U) [2 -
     3 \lambda_5 f(\rho_U)]^2  
     \nonumber
  \\ 
  &
  \left\{-6 \sqrt{6} \lambda_3 \lambda_5 f(\rho_U) -
     2\sqrt{2}[ \sqrt{3}  \lambda_3 + g(\rho_U)] +
     2 \sqrt{f(\rho_U)} [2  \sqrt{3}  \lambda_3 +
        g(\rho_U)] + \lambda_5 f^{3/2}(\rho_U) [11  \sqrt{3}  \lambda_3 + 2 
g(\rho_U)]\right\}
        \nonumber
        \\ 
        &+ 
  3  \frac{1}{2}  \lambda_3^2 f(\rho_U)
  [
     3 \lambda_5 f(\rho_U)-2]
     \left\{10  \sqrt{3}  \lambda_3 +
     6 g(\rho_U) -
     12 \sqrt{2} \lambda_5 \sqrt{
      f(\rho_U)} [ \sqrt{3}  \lambda_3 +
        g(\rho_U)] + \lambda_5 f(\rho_U) [21  \sqrt{3}  \lambda_3 + 19 
g(\rho_U)]
\right\}\Big\}~,
\label{prho}
\end{align}
\end{small}
where $f(\rho_U) =\frac{2\frac{1}{2} \rho_U + 6\lambda_5\rho_U^2 +
   \left[6 \lambda_3^2 \rho_U (\frac{1}{2} + 3 \lambda_5 
\rho_U)^2\right]^{1/2}}{(\frac{1}{2} +
   3 \lambda_5 \rho_U)^2} $ and $g(\rho_U) = \left[3 \lambda_3^2 + 4 \frac{1}{2} f(\rho_U) 
-
 6\frac{1}{2} \lambda_5 f^2(\rho_U)\right]^{1/2}$.
Hence, we can now calculate the total equation of state parameter
\begin{eqnarray}
 w_U(\rho_U)=\frac{p_U(\rho_U)}{\rho_U}
\end{eqnarray}
as a function  of the scale factor and of the coupling constants  that appear in
the considered action (\ref{actionfin}). Observing the form of expression $p_U(\rho_U)$ of
(\ref{prho}), we can see that there are
parameter regions which could give $p_U=0$ for a long time interval, that 
corresponds to a pressureless component, while departing from zero at late times,
tending asymptotically to the value $p_U(\rho_U)=-\rho_U$, i.e to $w_U=-1$.
In particular, we can
see that $p_U(\rho_U)$ may be considered as a generalization of the extended Chaplygin
gas
\cite{Bento:2002yx, Bento:2003we},
 however its specific form has arisen from some simple subclasses of the
Horndeski theory (this is different from the 
approach of  \cite{Kamenshchik:2001cp,Gorini:2005nw} and \cite{Granda:2011zy} in which 
the authors reconstructed  k-essence \cite{ArmendarizPicon:2000dh} models that may give 
rise to the simple Chaplygin gas). Note that in the shift-symmetric Kinetic Gravity 
Braiding models, without
$G_5$ and with $G_4$ as in general relativity, one effectively obtains  imperfect fluids, 
with zero vorticity and no dissipation  but with some diffusivity encoded in 
$G_3(X)$ \cite{Pujolas:2011he}, and thus we expect this description to be valid in the 
present work too.

To obtain a unified description of dark matter and dark energy in the above 
shift-symmetric generalized Galileon model we  solve equation
\eqref{frt} with respect to $X$ obtaining
\begin{equation}
X = \frac{3 H (\sqrt{2} \lambda_3+2 H)}{2 (\frac{1}{2} +9 \lambda_5 H^2)}~.
\label{Xauxilia}
\end{equation}
Then, substituting the above expression into the Klein-Gordon equation  \eqref{kgt} we 
get 
a simple
differential equation for $H(t)$, namely
\footnote
{  We mention here that in the case of shift-symmetric theories $P_\phi$ in 
(\ref{Pphidef}) 
is zero and then (\ref{fieldeq}) implies that $a^3 J=J_0=const.$. Thus, since during 
inflation $a$ increases 
exponentially, in late-time cosmology $J$ will be approximately zero, and since $J$ 
depends 
on $X$ and $H$ it will provide an algebraic equation $J(X, H)\approx0$ 
\cite{Deffayet:2010qz}. This  first integral  can be useful concerning the global features 
of the 
scenario, e.g. the extraction of dynamical attractors. However, in this work we are 
interested in the exact evolution at all times, and not only on the asymptotic behavior, 
and hence a parametric form of $H(X)$ is not adequate since we need both $H(z)$ and $X(z)$ 
in order to know the accurate evolution of various observables. Therefore, we need to 
indeed solve the differential equation  (\ref{eq1}).  Definitely, we can straightforwardly 
check that our solutions satisfy $a^3 
J=J_0=const.$, and we obtain a value of $J_0$ around 0.1 in units 
$8\pi G = c = \hbar = 1$ and with $H_0\approx 6 \times 10^{-61}$.  These large values of 
the shift-charge density  $J_0$ are only possible provided this effective fluid was formed 
after inflation or if there was some mechanism changing the shift-charge (this mechanism
would require a breaking of the shift-symmetry $\phi\rightarrow \phi+c$).
}
{\small{
\begin{equation}\label{eq1}
\!\!\!\!
\dot{H} =\frac{\sqrt{6} H^{3/2} (\frac{1}{2}\! +\!
   9 \lambda_5 H^2)\! \left\{
     \frac{1}{2}  (\sqrt{2}\lambda_3 \!+ \!2 H)^{3/2} (\frac{1}{2}\! +\! 9 \lambda_5 
H^2)^{1/2}\! -\!
   \sqrt{6H}  \left[\lambda_3^2 (\frac{1}{2}\! -\! 3 \lambda_5 H^2)\! +\!
      3 \sqrt{2} \lambda_3 H(\frac{1}{2} \!+\!\lambda_5 H^2) \!+\!
      4 H^2 (\frac{1}{2} + 3 \lambda_5 H^2)\right]\!\right\}}
      {
\lambda_3^2 (\frac{1}{2} - 27 \lambda_5 H^2) (\frac{1}{2} + 3\lambda_5 H^2) +
 6 \sqrt{2}\lambda_3 H (\frac{1}{2}^2 + 5 \frac{1}{2} \lambda_5 H^2 + 36\lambda_5^2 H^4) +
 8 H^2 \left[\frac{1}{2}^2 + 9\lambda_5 H^2 (\frac{1}{2}\! +\! 6\lambda_5 H^2)\right]}~.
\end{equation}}}
Finally, inserting (\ref{Xauxilia}) and (\ref{eq1}) into equation (\ref{EoS}) we obtain
  \begin{eqnarray}
\label{weq1}
&&
\!\!\!\!\!\!
w_U =
\Big\{3 H\left\{\lambda_3^2 (\frac{1}{2} - 27 \lambda_5 H^2) (\frac{1}{2} + 3\lambda_5 
H^2) +
   6\sqrt{2} \lambda_3 H (\frac{1}{2}^2 + 5 \frac{1}{2}\lambda_5 H^2 + 36 \lambda_5^2 H^4) 
+
   8 H^2 \left[\frac{1}{2}^2 + 9 \lambda_5 H^2 (\frac{1}{2} + 6 \lambda_5 
H^2)\right]\right\}
   \Big\}^{-1}
   \nonumber \\
&&
\ \ \ \ \
\cdot
\Big\{-2\sqrt{6Η} \frac{1}{2} (\sqrt{2} \lambda_3 + 2 H)^{3/2} (\frac{1}{2} + 9 \lambda_5 
H^2)^{3/2}
\nonumber \\
 && \ \ \ \ \,\ \ \ \
 +  3 H \left\{8 \frac{1}{2} H^2 (\frac{1}{2} + 15 \lambda_5 H^2) +
   3 \lambda_3^2 (\frac{1}{2}^2 + 16 \frac{1}{2} \lambda_5 H^2 - 9 \lambda_5^2 H^4) +
   6 \sqrt{2} \lambda_3 H \left[\frac{1}{2}^2 + 3 \lambda_5 H^2 (5 \frac{1}{2} - 6 
\lambda_5
H^2)\right]\right\}\Big\}~.
 \end{eqnarray}
 Hence, as long as we solve the differential equation  (\ref{eq1}), we have the solution
for the equation of state $w_U$ from equation (\ref{weq1}).

Equation  (\ref{eq1}) cannot be analytically solved in general. Therefore, in the
following we will proceed to numerical elaboration in order to extract the solution of
$H(t)$ and then the behaviour of $w_U$. For convenience, and in order
to compare our results with the observational data, we will use the redshift $z=-1+a_0/a$
as the independent variable, setting the current scale factor  $a_0$ to 1 (thus $\dot{H}
= -(1+z)H(z)H'(z)$ with prime denoting derivatives with respect to $z$).

\begin{figure}[ht]
\includegraphics[width=0.49 \linewidth]{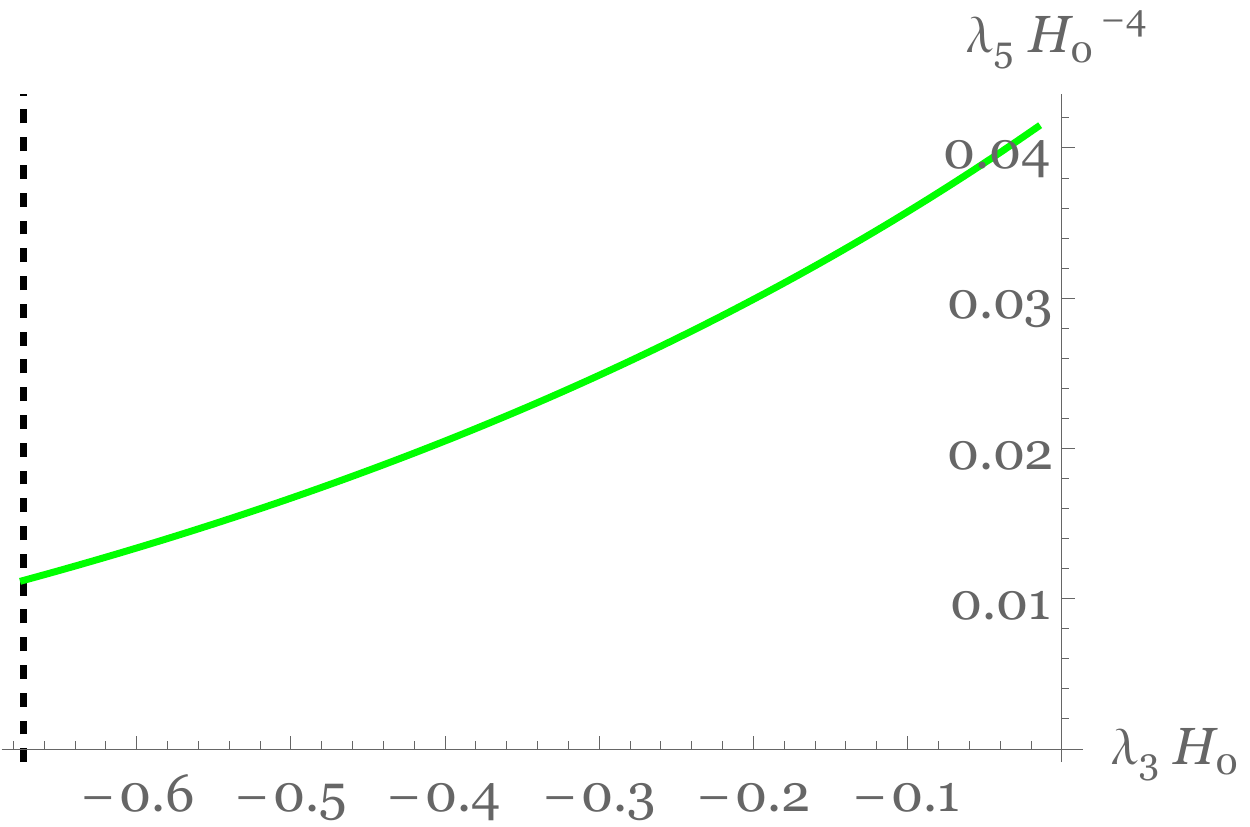}
\caption{{\it{\label{couplings}
The one-dimensional parameter subspace of the $(\lambda_3-\lambda_5)$ plane that 
satisfies the two phenomenological
requirements, namely  $H(z=0)\equiv H_0\approx 6 \times 10^{-61}$ (in units where $8\pi G 
= c = \hbar = 1$) and $w_U(z=0)\approx-0.7$. The dashed vertical line on the left marks 
the bound after which the two pathologies-absence conditions  \eqref{cscon} 
and \eqref{Qscon} are not satisfied.} }}
\end{figure}

We set the
present value (i.e at $z=0$) of $H$ to $H_0=\frac{h}{3000} Mpc^{-1}$, with the
dimensionless constant $h$ being around $0.69$, which in units where
$8\pi G = c = \hbar = 1$ used above reads  as  $H_0\approx 6 \times 10^{-61}$.
Additionally, we set the present value of the total equation of state of the Universe to
$w_U(z=0)\approx-0.7$, according to observations \cite{Ade:2015xua}. Hence,  the above 
two 
phenomenological
requirements are satisfied by an infinite number of pairs of the remaining parameters
$\lambda_3$ and $\lambda_5$, lying in a curve of the $(\lambda_3-\lambda_5)$ plane, shown 
in Fig.~\ref{couplings}.

\begin{figure}[!]
\begin{center}
\includegraphics[width=0.46 \linewidth]{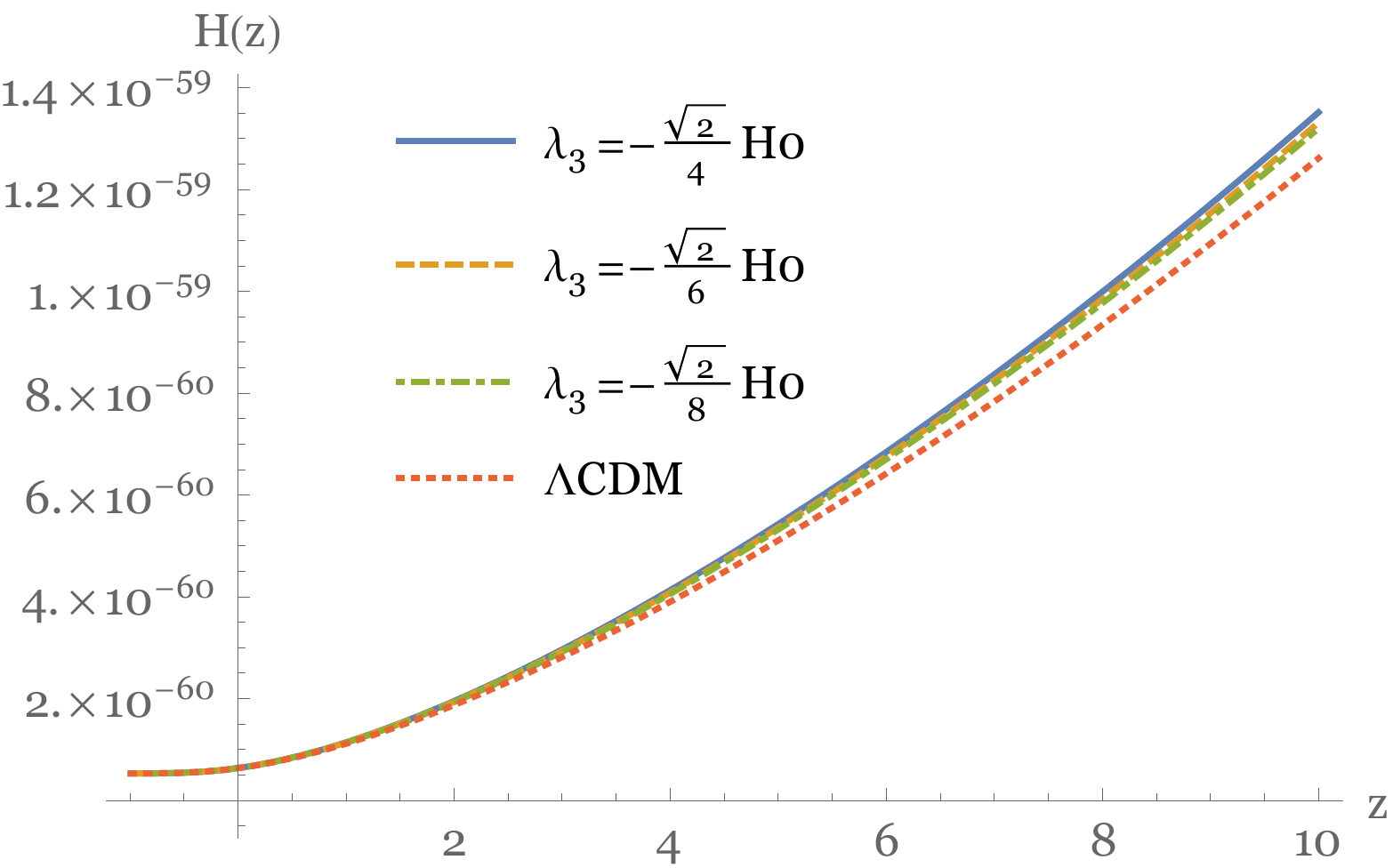}
\label{expansion_history}\quad\quad\quad
\includegraphics[width=0.46 \linewidth]{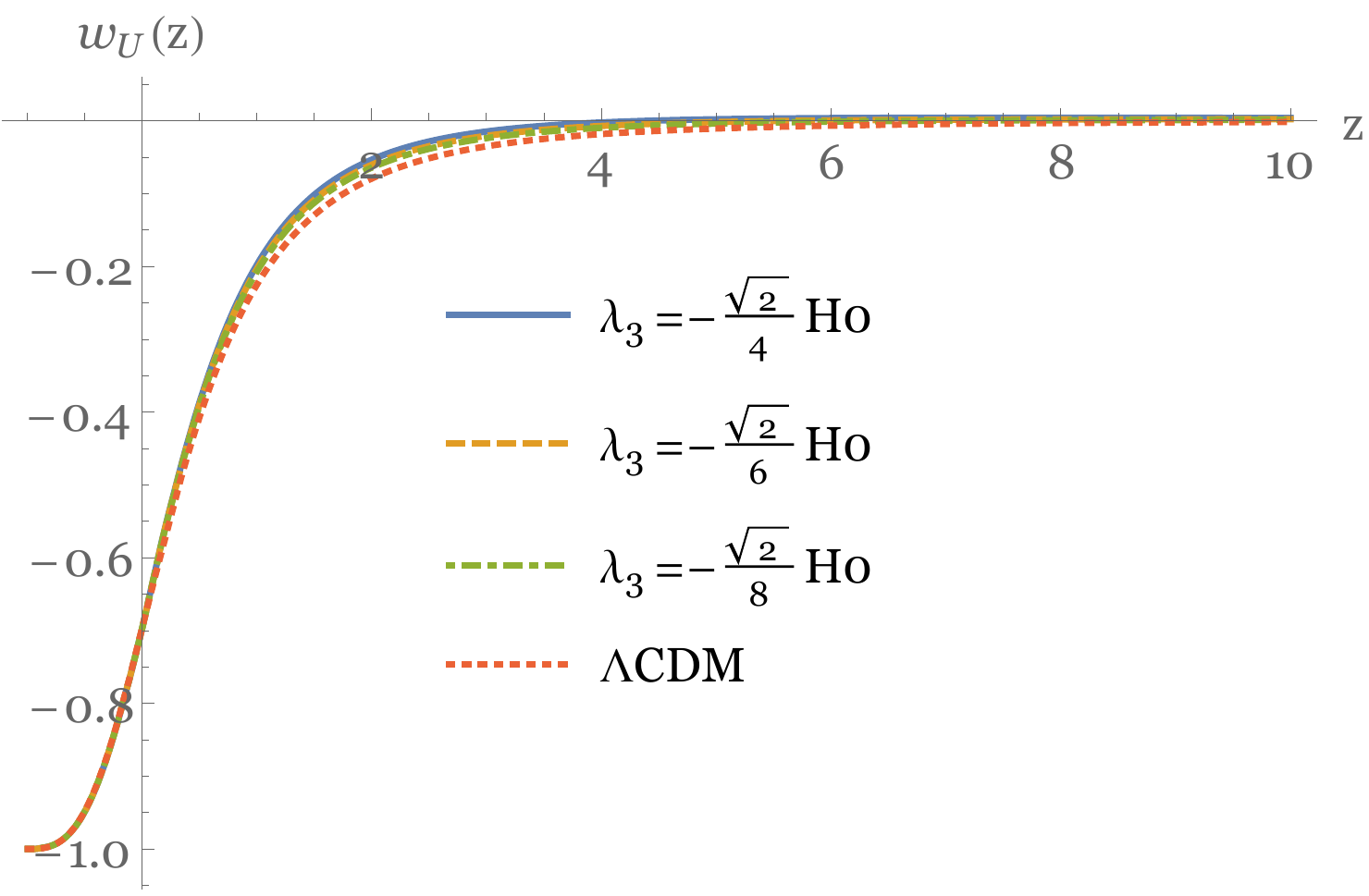}
\label{eos_U}
\caption{\it Left graph: The evolution of the Hubble parameter as a function of the
redshift $z$, for the model \eqref{actionfin}, for various values of $\lambda_3$ in units
where $8\pi G = c = \hbar = 1$. We have set the present value of the
total equation of state of the universe to $w_U(z=0)\approx-0.7$ and the present value of 
$H$ to
$H_0\approx 6 \times 10^{-61}$ which determines accordingly the value of $\lambda_5$ 
(see Fig.~\ref{couplings}). Right graph: 
The corresponding evolution of
$w_U(z)$. In both graphs we have added the corresponding curves of $\Lambda$CDM
paradigm.} \label{results}
\end{center}
\end{figure}

In the left graph of Fig. \ref{results} we present the solution for $H(z)$ from equation
(\ref{eq1}), for three choices of $\lambda_3$ and for comparison we
additionally depict the corresponding evolution from the $\Lambda$CDM model. In the right
graph of  Fig. \ref{results} we show the corresponding evolution for the total equation 
of 
state of the Universe from equation (\ref{weq1}), together with its evolution in the case 
of $\Lambda$CDM paradigm. 
Finally, for clarity in Fig.~\ref{logplot} we present $w_U(z)$ in logarithmic scale in 
order 
for the behaviour of larger $z$ to be visible.

\begin{figure}[ht!]
\begin{center}
\includegraphics[width=0.52 \linewidth]{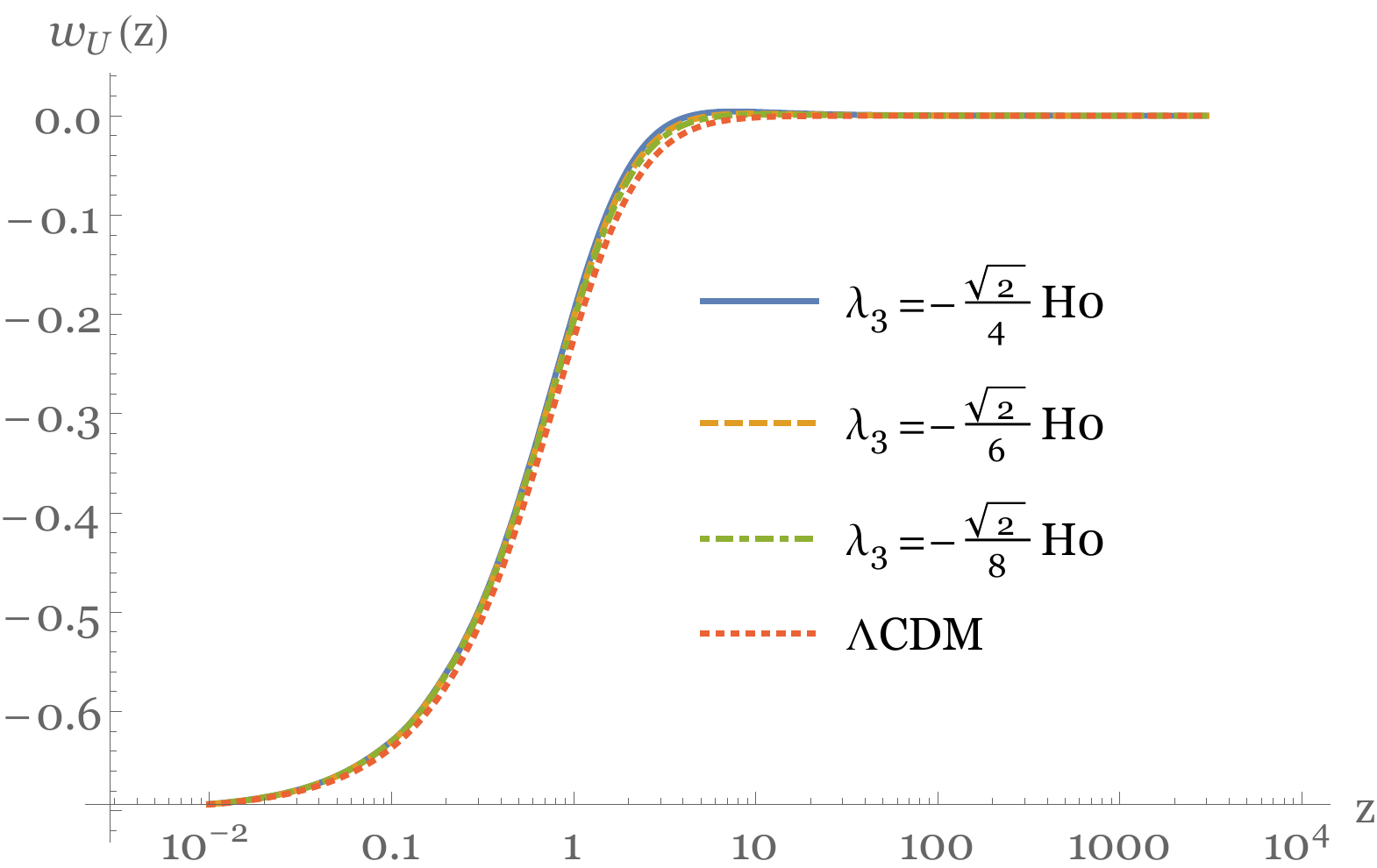}
\label{logplot}
\caption{\it  The evolution of 
$w_U(z)$ of the right graph of Fig.~\ref{results}, in logarithmic scale up to 
$z\sim3000$. } \label{logplot}
\end{center}
\end{figure}

As we observe the form of $w_U(z)$ is in excellent agreement with observations. It is zero
during the whole matter era, namely from  redshifts $z\sim3000$ up to $z\sim2-3$, then it
starts decreasing, passing the bound $-1/3$, which marks the onset of acceleration, at
around $z\sim0.5$, and finally it acquires a value $-0.7$ at present times (i.e at $z=0$).
We mention that this behaviour is obtained without considering an explicit matter
sector, namely it is the Galileon field itself that describes both the dark matter and
dark energy sectors in a unified way. Furthermore, for completeness, in the same figures
we have included the evolution up to the far future ($z\rightarrow-1$ corresponds to
$t\rightarrow\infty$), where we can clearly see that the universe tends toward a
de-Sitter phase. We stress that this de-Sitter phase is obtained although we have not
considered an explicit cosmological constant, and hence it is a very interesting 
non-trivial result.

One can acquire a qualitative picture of the above results in the following way. If we
manage to make $X$ to be almost constant at early times then we can expand $w_U$ from 
equation
(\ref{EoS}) for large $H$ (which is the case at early times) yielding an approximated
value $w_U \rightarrow 0$. On the other hand, if $X$ departs from the constant
value, acquiring a decreasing form, then $w_U$ starts deviating from 0 towards the  $-1$
value. Our model  (\ref{actionfin}) presents these behaviours, and indeed one can see that
for large $H$ we obtain $X \rightarrow \frac{1}{3\lambda_5} = \text{constant}$  as
requested.
On the other hand, if we had included an extended non-minimal derivative coupling $\sim
G_5(X)$ \cite{Harko:2016xip}, then the introduced extra  $\ddot{\phi}$ term would have 
changed the 
kinetic energy of the cosmological fluid spoiling  the   $w_U \rightarrow
0$ feature at early times, leading instead to $w_U \rightarrow
const.$ However in that case, adding suitably more terms in the action  one could realize 
the very 
interesting scenario that $w_U$ starts from $-1$  at very early times, then it departs 
towards $1/3$ remaining there for a sufficient time interval, then becoming zero for a 
very long time and 
finally departing towards $-1$ at late and future times. Such an evolution could describe 
the whole 
thermal history of the Universe, namely the inflationary, the radiation, the matter and 
finally the 
dark energy eras. 

%

 Let us now confront the model at hand with cosmological observations coming from
Supernovae type Ia (SN Ia), which is straightforward as long as we have the evolution of
$H(z)$.
Observations measure the apparent luminosity vs redshift $l(z)$, or equivalently the 
apparent 
magnitude vs redshift $m(z)$, which are related to the luminosity distance by
\be
2.5 \log\left[\frac{L}{l(z)}\right] = \mu \equiv m(z) - M = 5 
\log\left[\frac{d_L(z)_{\text{obs}}}{Mpc}\right]  + 25
~,
\ee
where $L$ and $M$ are respectively the absolute luminosity and magnitude.
From the theory point of view the predicted Hubble parameter is related to the 
dimensionless 
luminosity distance $d_{L}(z)_\text{th}$ as
\begin{equation}
d_{L}\left(z\right)_\text{th}\equiv\left(1+z\right)
\int^{z}_{0}\frac{dz'}{H\left(z'\right)}~.
\end{equation}

\begin{figure}[ht]
\centering
\includegraphics[scale=0.60]{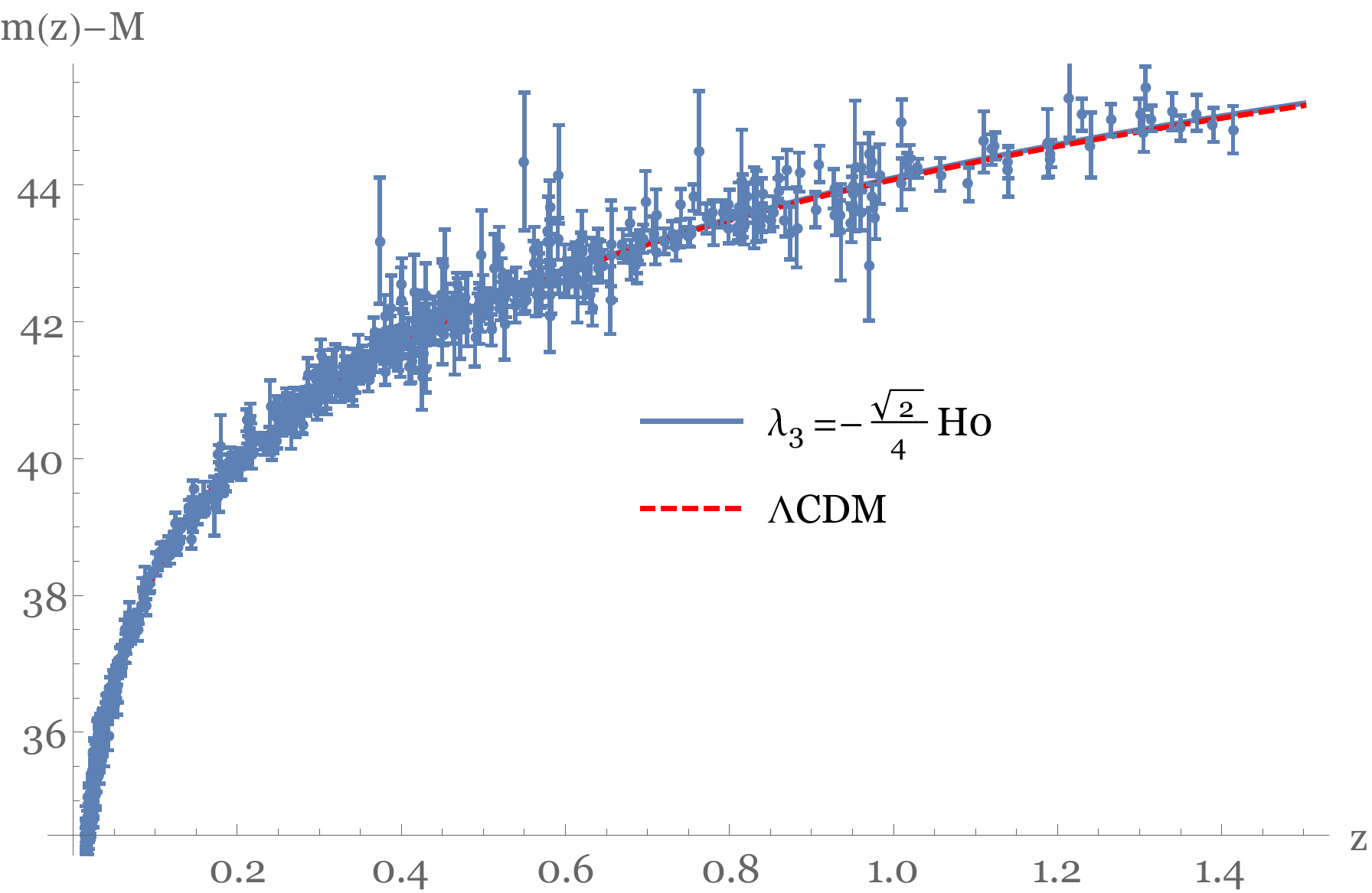}
\caption{ {\it{Theoretically predicted apparent minus absolute magnitude
for the model (\ref{actionfin}), for  $\lambda_3=-\sqrt{2}H_0/4$ in units
where $8\pi G = c = \hbar = 1$, on top of the $580$ SN Ia data points from
\cite{Suzuki:2011hu}.
For comparison, we have added the corresponding curve of $\Lambda$CDM paradigm.
 }}}
\label{LCDM_fit}
\end{figure}

In our model the evolution of $H(z)$ is given numerically from the
differential equation  (\ref{eq1}), and it was presented in the left graph of
Fig. \ref{results}. On the other hand, for the case of $\Lambda$CDM paradigm, the
corresponding  $H(z)$ is given by $ H^2(z) = H_0^2\left[\Omega_{m0}(1+z)^3 +
\Omega_{\Lambda0} \right]$, with $\Omega_m$ and $\Omega_\Lambda$ the matter and
cosmological-constant density parameters respectively, and where the subscript ``0''
denotes the present value of these quantities. In Fig.   \ref{LCDM_fit} we present
the theoretically predicted apparent minus absolute magnitude
for the model (\ref{actionfin}), on top of the $580$ SN Ia data points from
\cite{Suzuki:2011hu},
as well as the corresponding curve from the $\Lambda$CDM scenario. As we observe, the 
agreement
with the SN Ia data is excellent. The proposed model  of dark matter - dark energy 
unification
is almost indistinguishable from  $\Lambda$CDM cosmology, although we have not considered
an explicit cosmological constant and an explicit matter sector. This is one of the main
results of the present work.

A comparison of the fittings of the above two scenarios can be achieved using the Akaike 
Information
 Criterion $(AIC)$ \cite{Akaike1974}. The $AIC$ is defined as
\be
AIC = -2\ln\mathcal{L} + 2 d = \chi_{min}^2 + 2d,
\ee
with $\mathcal{L} = \exp\left(-\chi^2_{min}/2\right)$ the maximum likelihood function and  
$d$ the 
number of model parameters. The minimization of $\chi^2$ is defined as
\be 
\chi^2_{min}= \sum^N_i \frac{\left[\mu(z_i)_{obs} - \mu(z_i)_{th}\right]^2 }{\sigma_i^2},
\ee
with $N=580$ the number of SN Ia contained in \cite{Suzuki:2011hu} and $\sigma_i$ the 
error 
corresponding to each $\mu(z_i)_{obs}$. The $\Lambda$CDM cosmology will be the reference 
scenario, 
with respect of which the comparison of the model under consideration will be performed. 
For any 
specific model $M$, calculating the difference $\Delta AIC = {AIC}_M - 
{AIC}_{\Lambda\text{CDM}}$ 
and resulting to a value $\leq 2$, the considered model has substantial observational 
support with 
respect to the $\Lambda$CDM one  \cite{Nunes:2016drj,delaCruz-Dombriz:2016bqh}. In the 
above 
analysis we have one parameter for $\Lambda$CDM paradigm, while in our model we have three 
free 
parameters. This results to a value 
\be 
\Delta AIC = 0.65 < 2,
\ee
and therefore the scenario at hand is very efficient concerning confrontation with 
observations.

To complete our analysis on the above  scenario that describes in a unified way the dark 
matter and dark energy sectors, we must examine whether the model is free from 
pathologies 
such as ghosts and Laplacian instabilities at the perturbative level. Indeed, it is well 
known that in many modified gravities there appear various pathologies at the 
perturbative level, even if the background behaviour is problem-free and consistent. In 
the Horndeski theory, the two necessary conditions for absence of pathologies in the 
scalar perturbations are given 
in 
relations \eqref{cscon} and \eqref{Qscon}, which in the particular model of action 
(\ref{actionfin}) become
\begin{equation}
\!\!\!\!\!\!\!\!\!\!\!\!\!\!\!\!
Q_s = - \frac{2 ( \lambda_5 X -1) \Big\{3 \lambda_3^2 +
    6 \sqrt{2}\lambda_3 H  (1 - 5 \lambda_5 X ) +
    8 X  \left[\frac{1}{2} - \frac{1}{2} \lambda_5 X  +
       3 \lambda_5 H^2  (1 + 3 \lambda_5 X )\right]\Big\}}
       {\left[\sqrt{2} \lambda_3 + H  (4 - 12
\lambda_5 X)\right]^2} > 0~,
\label{Qsrelation}
\end{equation}
\begin{eqnarray}
&&
\!\!\!\!\!\!\!\!\!\!\!\!\!\!
c_s^2  =
 \Big\{( \lambda_5 X -1) \left\{3
\lambda_3^2 +
   6 \sqrt{2}  \lambda_3 H  (1 - 5 \lambda_5 X ) +
   8 X  \left[\frac{1}{2} - \frac{1}{2} \lambda_5 X  +
      3 \lambda_5 H^2  (1 + 3\lambda_5 X )\right]\right\}\Big\}^{-1}\nonumber\\
      &&
      \cdot
\Big\{32 \lambda_5^2 H^2  X^2  ( 3\lambda_5 X -1) -
 8 ( \lambda_5 X  - 1)^2 ( 3 \lambda_5 X -1 ) \dot{H} + \lambda_3 \left[\lambda_3 +
\lambda_3 \lambda_5 X  -
   4 \sqrt{2} \lambda_5 (\lambda_5 X -1) \dot{X} \right]\nonumber
      \\
      &&
  \ \ \
  -2 H  \left\{\sqrt{2} \lambda_3 \left[\lambda_5 X  (2 + 7 \lambda_5  X
)-1\right] -
   4 \lambda_5  ( \lambda_5  X -1) (1 +
      3 \lambda_5  X ) \dot{X} \right\}
       \Big\} \geq 0~,
       \label{cs2relation}
\end{eqnarray}
with the additional constraint $c_s^2 \leq 1$. These conditions must be valid on-shell of
the background evolution and at all times.

\begin{figure}[!]
\centering
\includegraphics[width=0.46 \linewidth]{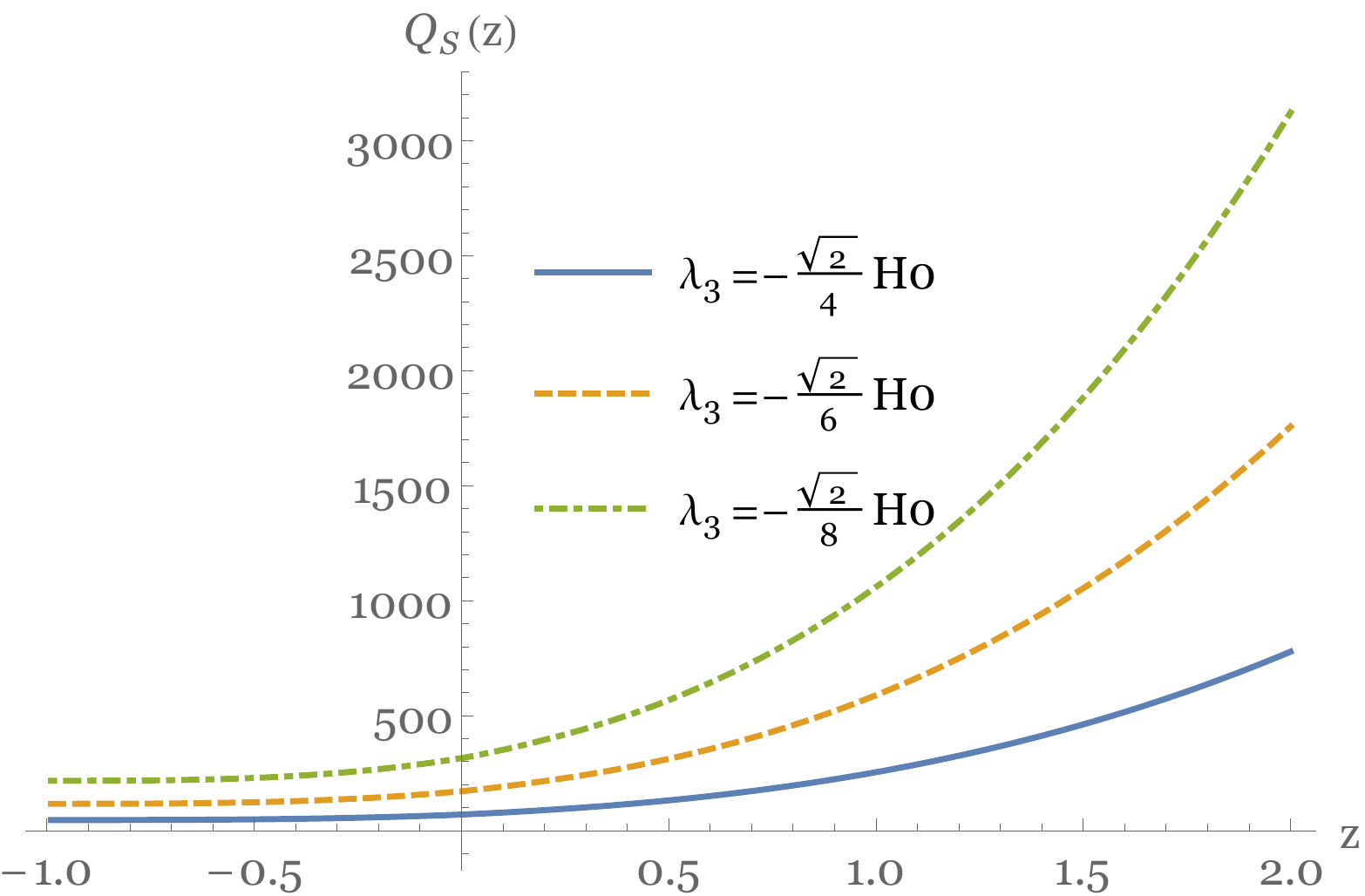}
\label{sound_speed}\quad\quad\quad
\includegraphics[width=0.46 \linewidth]{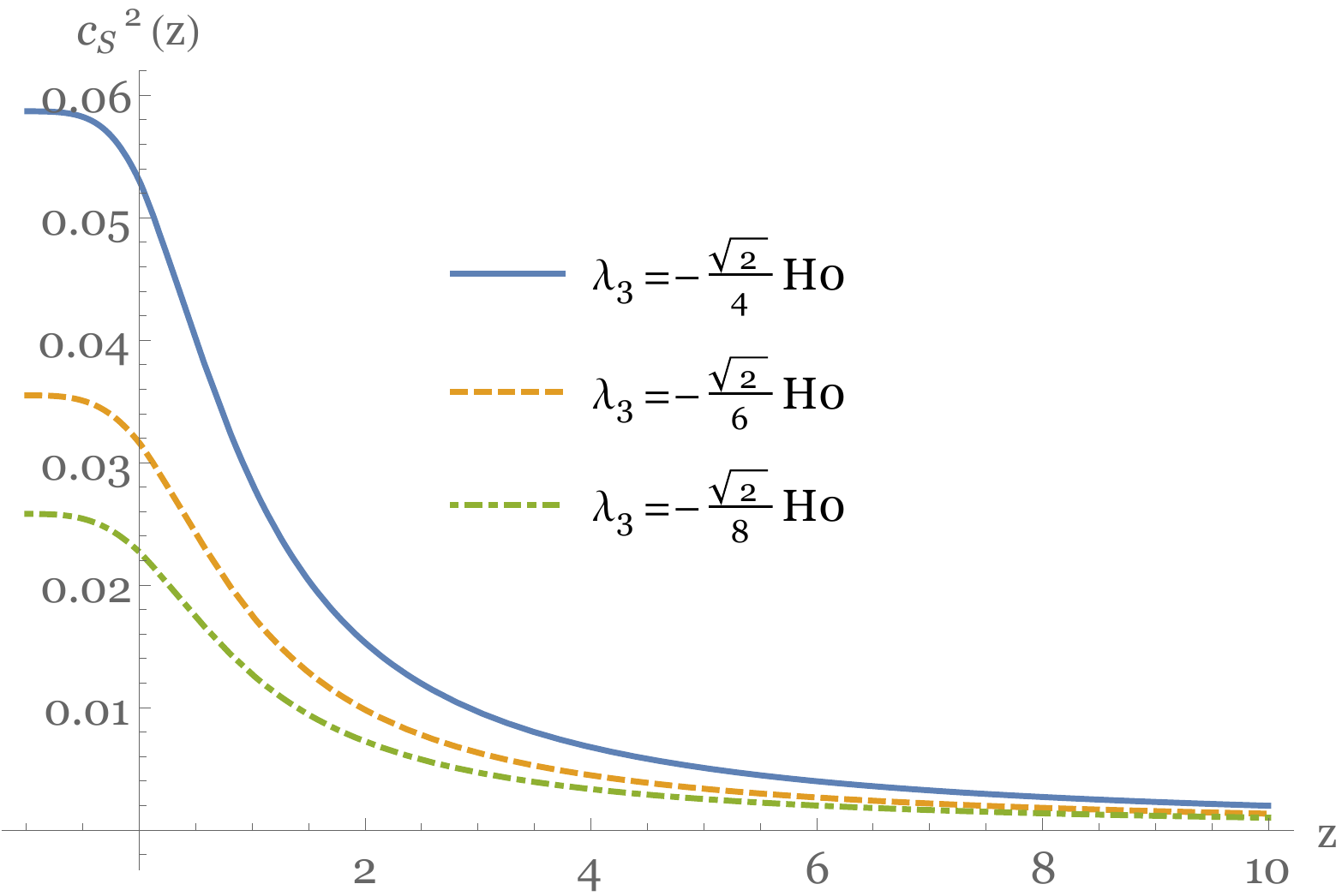}
\label{ghost_free}
\caption{ \it Left graph: The evolution of $Q_s$ from (\ref{Qsrelation}), that
characterizes the absence of Laplacian instabilities, as a function of the
redshift $z$, for the model (\ref{actionfin}), for various values of $\lambda_3$ in units
where $8\pi G = c = \hbar = 1$. We have set the present value of the
total equation of state of the universe to $w_U(z=0)\approx-0.7$ and the present value of 
$H$ to
$H_0\approx 6 \times 10^{-61}$. Right graph: The corresponding evolution of
the sound speed square of the scalar
perturbations $c_s^2$ from (\ref{cs2relation}).  } \label{pertu}
\end{figure}

In Fig. \ref{pertu} we present the evolution of  $Q_s$ and $c_s^2$ for the background
evolution and parameter choices of Fig. \ref{results}. We observe that the conditions for
absence of ghosts and Laplacian instabilities are always satisfied, and thus the
scenario at hand is free from pathologies of the scalar sector at both background and 
perturbative level.
We mention here that although the desired background evolution can be relatively easily
obtained with an action of the form (\ref{action1}), to satisfy  the
pathologies-absence conditions requires careful selection of the involved functions.
Hence, as we mentioned earlier, in the particular model (\ref{actionfin}) the choice of
$X^{1/2}$ in the specific forms of  $K(\phi,X)$ and  $G_3(\phi,X)$ is not
necessary for obtaining the desired phenomenological evolution at the
background level, however it is important in order to make  $c_s^2$ positive at all 
times, since the Cuscuton and the Kinetic 
Gravity Braiding terms   (which exhibit 
superluminal sound speed) ensure that the sound speed square will increase and avoid 
negative values.
\begin{figure}[!]
\centering
\includegraphics[width=0.46 \linewidth]{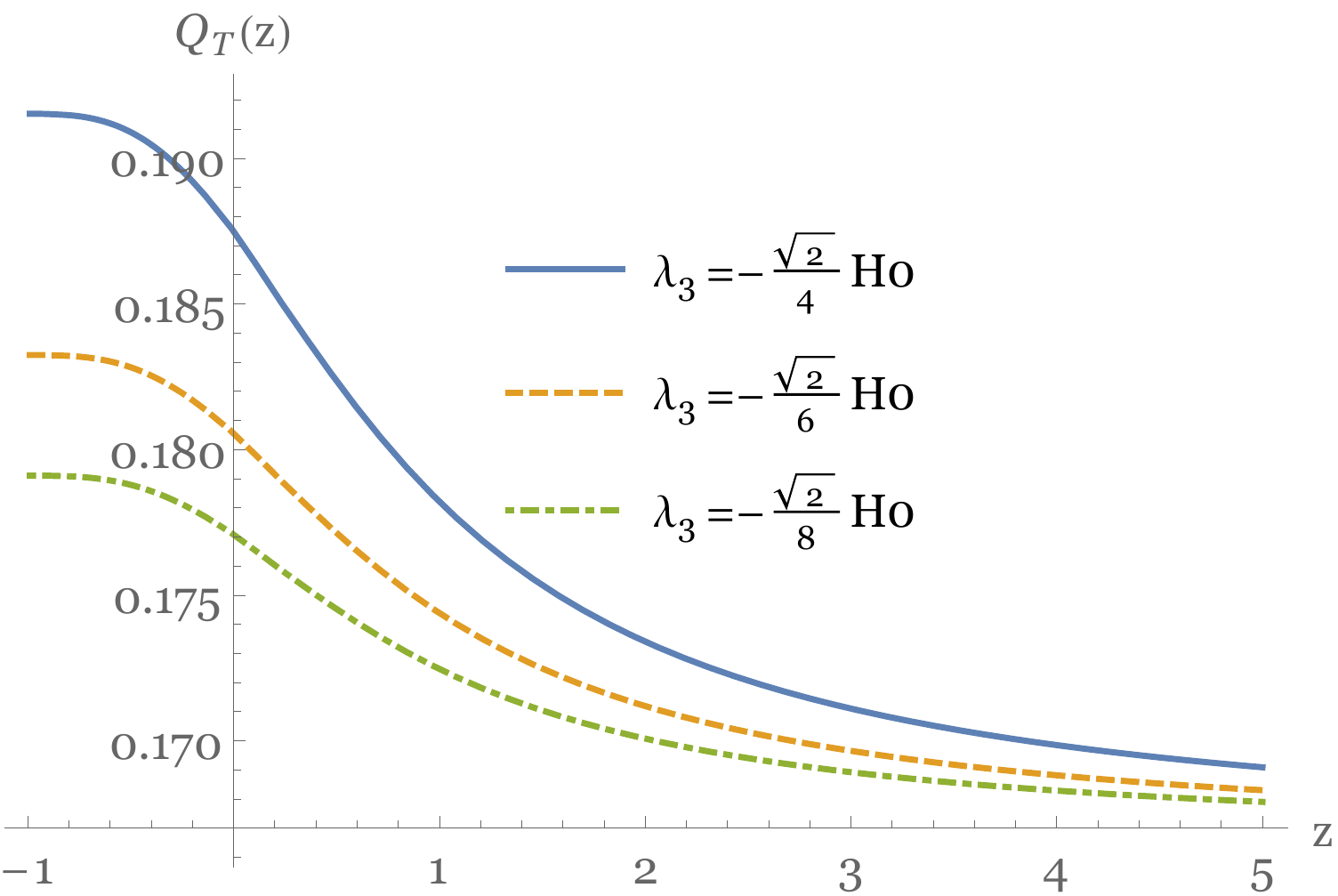}
\quad\quad\quad
\includegraphics[width=0.46 \linewidth]{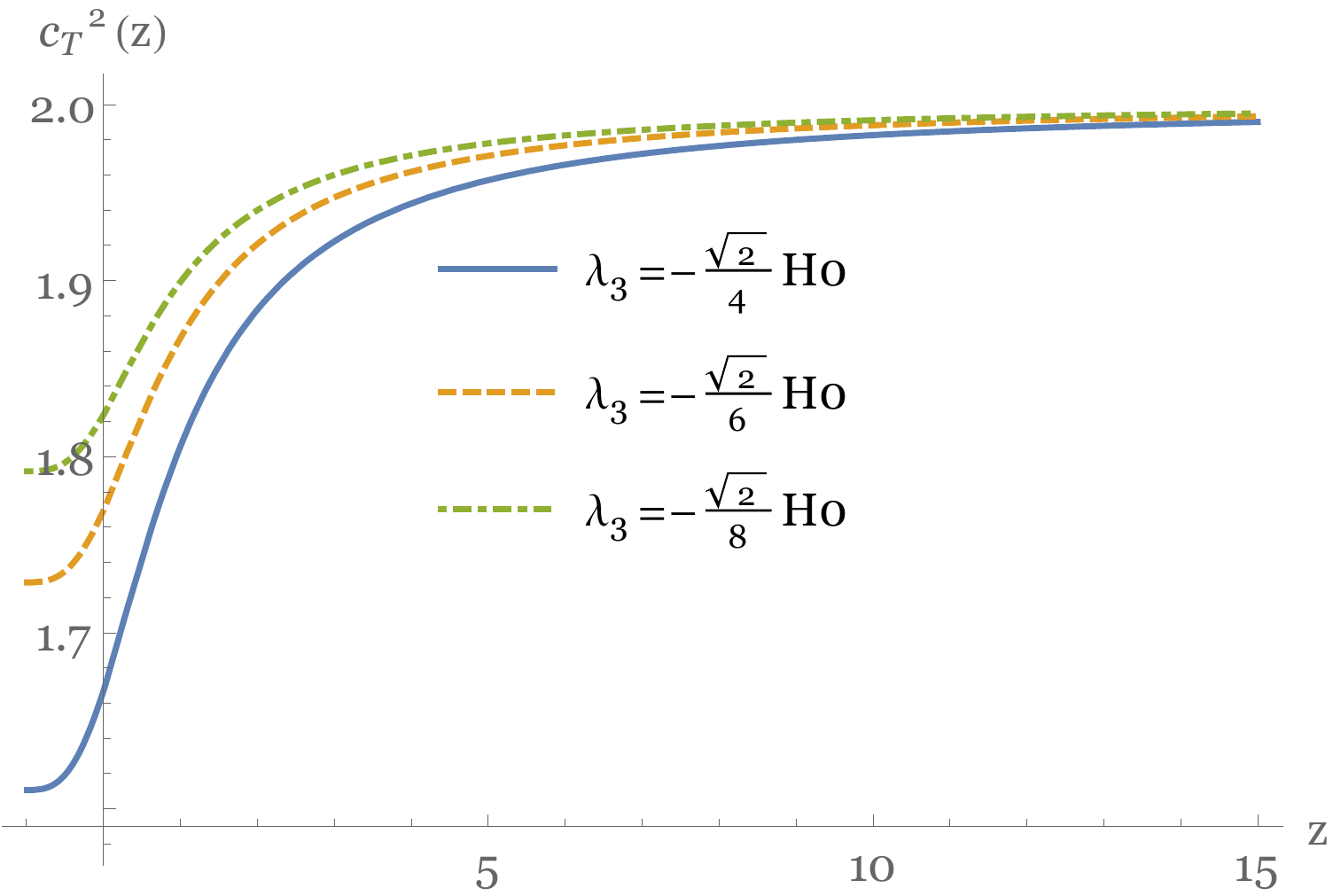}
\caption{ \it Left graph: The evolution of $Q_T$ from (\ref{qtpert}), that
characterizes the absence of Laplacian instabilities, as a function of the
redshift $z$, for the model (\ref{actionfin}), for various values of $\lambda_3$ in units
where $8\pi G = c = \hbar = 1$. We have set the present value of the
total equation of state of the universe to $w_U(z=0)\approx-0.7$ and the present value of 
$H$ to
$H_0\approx 6 \times 10^{-61}$. Right graph: The corresponding evolution of
the sound speed square of the tensor
perturbations $c_T^2$ from (\ref{ctsquare}).  } \label{tpertu}
\end{figure}

Finally, concerning tensor perturbations, the two necessary conditions for 
absence of ghost and Laplacian instabilities are given in  \eqref{tens1} and 
\eqref{tens2}, and thus in the particular model of action 
(\ref{actionfin}) become
\be 
\label{qtpert}
Q_T = \frac{1}{4} (1-\lambda_5 X)
\ee
\be  \label{ctsquare}
c_T^2 = \frac{\lambda_5 X+1}{1-\lambda_5 X} \geq 0 .
\ee
 In Fig.~\ref{tpertu} we depict the evolution of  $Q_T$ and $c_T^2$ for the 
background evolution and parameter choices of Fig. \ref{results}. As we can see,  the 
conditions for absence of ghosts and Laplacian instabilities are always satisfied, and 
thus the scenario at hand is free from pathologies in the tensor sector too. 
Additionally, note that since the specific model considered here contains  $G_5$, the 
tensor perturbations may become superluminal,  as expected \cite{DeFelice:2011bh, 
Germani:2010ux}. However, as discussed above, this feature does not imply pathologies or 
acausality  \cite{Babichev:2007dw,Deffayet:2010qz}.  \footnote{
 While this manuscript was in the revision process, the LIGO-VIRGO collaboration 
detected a binary neutron star merger with gravitational waves
(GW170817) \cite{TheLIGOScientific:2017qsa}, and the Fermi Gamma-ray Burst
Monitor detected  its associated electromagnetic counterparts 
\cite{Monitor:2017mdv},
which imposes strong 
constraints on the gravitational waves speed ($|c_T/c-1|\leq 4.5\times 10^{-16}$    
\cite{Ezquiaga:2017ekz,Baker:2017hug}). Hence, one should try to focus on solutions where 
$X$ (i.e. 
$\dot{\phi}$) becomes very small at least in the late-time universe, in order for $c_T$ 
in (\ref{ctsquare}) to enter into the above bounds. The subtle and detailed elaboration 
of these solution subclasses is left for a separate project.}

\section{Conclusions}
\label{conclusions}

In this work we have presented a unified description of the dark matter and dark energy
sectors, in the framework of shift-symmetric generalized Galileon theories. Although in
usual applications of Galileon/Horndeski theories one includes the extra scalar degree
of freedom in order to represent the inflaton or the dark-energy field, while the matter 
sector is considered additionally, in our approach we used this scalar field in order to 
account for both dark matter and dark energy sectors simultaneously. Using a specific 
combination of terms 
of the  Galileon/Horndeski Lagrangian, and without considering an explicit matter sector 
and an explicit cosmological constant, we  obtained an effective cosmic fluid that
behaves as pressureless matter during the matter era, and as dark energy at late times,
thus producing a unified description of the cosmological evolution.

In particular, the shift symmetry  imposed on the action allowed the derivation of the 
pressure
of the effective fluid as a function of the energy density, namely
$p_U=p_U(\rho_U)$, which proves to have the form of an extended Chaplygin gas. Chaplygin
gas has been known to be efficient in offering a unified description of the matter and 
dark energy regimes. In the present model the
generalized equation of state parameter for the unified fluid $w_U$ is acquired
straightaway from the action considered.

The obtained behaviour of $w_U(z)$ is in excellent agreement with observations. It is zero
during the whole matter era, namely from  redshifts $z\sim3000$ up to $z\sim2-3$, then it
starts decreasing, passing the bound $-1/3$, which marks the onset of acceleration, at
around $z\sim0.5$, and finally it acquires a value $-0.7$ at present times. Additionally,
confronting this evolution with the Supernovae type Ia data we showed that we
obtained an excellent fit, and our model of dark-matter - dark-energy unification
is almost indistinguishable from  $\Lambda$CDM cosmology. Moreover, investigating the
evolution up to the far future we saw that the universe tends toward a de-Sitter phase.
We stress that these behaviours are obtained without considering an explicit matter
sector or an explicit cosmological constant, namely it is the Galileon field itself that
describes both the dark matter and dark energy sectors in a unified way. This is the main
result of the present work.

Finally, we examined in detail the behaviour of the above unified scenario at the 
perturbative level, focusing on the conditions of absence of pathologies such as ghosts 
and Laplacian instabilities, at both scalar and tensor sectors. This is a crucial 
step that must be taken in every 
cosmological application, since it is well known that in many modified theories of
gravities there appear various pathologies at the perturbative level, even if the
background behaviour is problem-free and consistent. We showed that with the consideration
of specific terms in the action  the various pathologies-absence
conditions are satisfied at all times on top of the background unified solutions.

As a further work on this unification approach of the dark sectors it would be 
interesting 
to consider the powerful method of dynamical system analysis in order to bypass the 
nonlinearities of the equations and extract the global features and the asymptotic 
behaviour of the evolution. Furthermore, apart from the SN Ia data, one should use 
observations from Baryon Acoustic Oscillations (BAO), from Cosmic Microwave Background 
(CMB), and from Large Scale Structure (LSS), in order to obtain a more detailed and 
complete confrontation, since this was the weak point of previously constructed unified 
scenarios such as the Chaplygin gas.
Additionally, one should try to focus on solutions where the scalar-field kinetic energy  
becomes very small, at least in the late-time universe, in order for the tensor 
perturbation speed to get reduced close to the light speed.
In order to pass these necessary observational
comparisons one must consider the baryonic matter sector explicitly, since this will not
affect the obtained unified description of the dark matter and dark energy sectors.




\begin{thebibliography}{99}

\bibitem{acceleration}
  S.~Perlmutter {\it et al.} [Supernova Cosmology Project Collaboration],
  ``Discovery of a supernova explosion at half the age of the Universe and its 
cosmological implications,''
  Nature {\bf 391}, 51 (1998)
  [astro-ph/9712212].


\bibitem{ries}
 A.~G.~Riess {\it et al.} [Supernova Search Team],
  ``Observational evidence from supernovae for an accelerating universe and a 
cosmological constant,''
  Astron.\ J.\  {\bf 116}, 1009 (1998)
  [astro-ph/9805201].
 
\bibitem{Copeland:2006wr} 
  E.~J.~Copeland, M.~Sami and S.~Tsujikawa,
  ``Dynamics of dark energy,''
  Int.\ J.\ Mod.\ Phys.\ D {\bf 15}, 1753 (2006)
  [hep-th/0603057].

\bibitem{Joyce:2014kja}
  A.~Joyce, B.~Jain, J.~Khoury and M.~Trodden,
  ``Beyond the Cosmological Standard Model,''
  Phys.\ Rept.\  {\bf 568}, 1 (2015)
  [arXiv:1407.0059 [astro-ph.CO]].




\bibitem{Wang:2016lxa}
  B.~Wang, E.~Abdalla, F.~Atrio-Barandela and D.~Pavon,
  ``Dark Matter and Dark Energy Interactions: Theoretical Challenges, Cosmological
Implications and
Observational Signatures,''
  Rept.\ Prog.\ Phys.\  {\bf 79}, no. 9, 096901 (2016)
  [arXiv:1603.08299 [astro-ph.CO]].

\bibitem{Bolotin:2013jpa}
  Y.~L.~Bolotin, A.~Kostenko, O.~A.~Lemets and D.~A.~Yerokhin,
  ``Cosmological Evolution With Interaction Between Dark Energy And Dark Matter,''
  Int.\ J.\ Mod.\ Phys.\ D {\bf 24}, no. 03, 1530007 (2014)
  [arXiv:1310.0085 [astro-ph.CO]].

\bibitem{Bento:2002ps}
  M.~C.~Bento, O.~Bertolami and A.~A.~Sen,
  ``Generalized Chaplygin gas, accelerated expansion and dark energy matter
  Phys.\ Rev.\ D {\bf 66}, 043507 (2002).
  [gr-qc/0202064].

\bibitem{Bento:2002yx}
  M.~d.~C.~Bento, O.~Bertolami and A.~A.~Sen,
  ``Generalized Chaplygin gas and CMBR constraints,''
  Phys.\ Rev.\ D {\bf 67}, 063003 (2003)
  [astro-ph/0210468].

\bibitem{Bento:2005un}
  M.~C.~Bento, O.~Bertolami, M.~J.~Reboucas and P.~T.~Silva,
  ``Generalized Chaplygin gas model, supernovae and cosmic topology,''
  Phys.\ Rev.\ D {\bf 73}, 043504 (2006)
  [gr-qc/0512158].

  
  
\bibitem{Sandvik:2002jz} 
  H.~Sandvik, M.~Tegmark, M.~Zaldarriaga and I.~Waga,
  ``The end of unified dark matter?,''
  Phys.\ Rev.\ D {\bf 69}, 123524 (2004)
  [astro-ph/0212114].

  
  
\bibitem{Farooq:2010xm}
  M.~U.~Farooq, M.~Jamil and M.~A.~Rashid,
  ``Interacting entropy-corrected holographic Chaplygin gas model,''
  Int.\ J.\ Theor.\ Phys.\  {\bf 49}, 2334 (2010)
  [arXiv:1003.3399 [gr-qc]].

\bibitem{Gorini:2007ta}
  V.~Gorini, A.~Y.~Kamenshchik, U.~Moschella, O.~F.~Piattella and A.~A.~Starobinsky,
  ``Gauge-invariant analysis of perturbations in Chaplygin gas unified models of dark 
matter and 
dark energy,''
  JCAP {\bf 0802}, 016 (2008)
  [arXiv:0711.4242 [astro-ph]].

\bibitem{Debnath:2004cd}
  U.~Debnath, A.~Banerjee and S.~Chakraborty,
  ``Role of modified Chaplygin gas in accelerated universe,''
  Class.\ Quant.\ Grav.\  {\bf 21}, 5609 (2004)
  [gr-qc/0411015].

\bibitem{BouhmadiLopez:2004mp}
  M.~Bouhmadi-Lopez and P.~Vargas Moniz,
  ``FRW quantum cosmology with a generalized Chaplygin gas,''
  Phys.\ Rev.\ D {\bf 71}, 063521 (2005)
  [gr-qc/0404111].

\bibitem{Setare:2007mp}
  M.~R.~Setare,
  ``Interacting holographic generalized Chaplygin gas model,''
  Phys.\ Lett.\ B {\bf 654}, 1 (2007)
  [arXiv:0708.0118 [hep-th]].



\bibitem{Fujii:2003pa}
  Y.~Fujii and K.~Maeda,
{\it{The scalar-tensor theory of gravitation}},
   Cambridge University Press, Cambridge (2003).


\bibitem{Horndeski:1974wa}
  G.~W.~Horndeski,
  ``Second-order scalar-tensor field equations in a four-dimensional space,''
  Int.\ J.\ Theor.\ Phys.\  {\bf 10} (1974) 363.

\bibitem{Ostrogradsky:1850fid}
  M.~Ostrogradsky,
  ``Mémoires sur les équations différentielles, relatives au problème des
isopérimètres'',
  Mem.\ Acad.\ St.\ Petersbourg {\bf 6}, no. 4, 385 (1850).

\bibitem{Nicolis:2008in}
  A.~Nicolis, R.~Rattazzi, E.~Trincherini,
  ``The Galileon as a local modification of gravity,''
  Phys.\ Rev.\  {\bf D79 } (2009)  064036
  [arXiv:0811.2197 [hep-th]].

\bibitem{Deffayet:2009wt}
  C.~Deffayet, G.~Esposito-Farese, A.~Vikman,
  ``Covariant Galileon,''
  Phys.\ Rev.\  {\bf D79 } (2009)  084003
  [arXiv:0901.1314 [hep-th]].

\bibitem{Deffayet:2009mn}
  C.~Deffayet, S.~Deser and G.~Esposito-Farese,
  ``Generalized Galileons: All scalar models whose curved background extensions
  maintain second-order field equations and stress-tensors,''
  Phys.\ Rev.\  D {\bf 80}, 064015 (2009)
  [arXiv:0906.1967].

\bibitem{Deffayet:2011gz}
  C.~Deffayet, X.~Gao, D.~A.~Steer and G.~Zahariade,
  ``From k-essence to generalised Galileons,''
  Phys.\ Rev.\ D {\bf 84} (2011) 064039
  [arXiv:1103.3260 [hep-th]].

\bibitem{Vainshtein:1972sx}
  A.~I.~Vainshtein,
  ``To the problem of nonvanishing gravitation mass,''
   Phys.\ Lett.\ B {\bf 39}, 393 (1972).



\bibitem{Chow:2009fm}
  N.~Chow and J.~Khoury,
  ``Galileon Cosmology,''
  Phys.\ Rev.\ D {\bf 80}, 024037 (2009)
[arXiv:0905.1325]].

\bibitem{DeFelice:2011th}
  A.~De Felice, R.~Kase and S.~Tsujikawa,
  ``Vainshtein mechanism in second-order scalar-tensor theories,''
   Phys.\ Rev.\ D {\bf 85}, 044059 (2012)
[arXiv:1111.5090]].


\bibitem{Babichev:2011kq}
  E.~Babichev, C.~Deffayet and G.~Esposito-Farese,
  ``Improving relativistic MOND with Galileon k-mouflage,''
   Phys.\ Rev.\ D {\bf 84}, 061502 (2011)
[arXiv:1106.2538]].


  
\bibitem{Saridakis:2016ahq} 
  E.~N.~Saridakis and M.~Tsoukalas,
  ``Cosmology in new gravitational scalar-tensor theories,''
  Phys.\ Rev.\ D {\bf 93}, no. 12, 124032 (2016)
  [arXiv:1601.06734 [gr-qc]].
  
  
\bibitem{deRham:2011by} 
  C.~de Rham and L.~Heisenberg,
  ``Cosmology of the Galileon from Massive Gravity,''
  Phys.\ Rev.\ D {\bf 84}, 043503 (2011)
  [arXiv:1106.3312 [hep-th]].
  
\bibitem{Heisenberg:2014kea} 
  L.~Heisenberg, R.~Kimura and K.~Yamamoto,
  ``Cosmology of the proxy theory to massive gravity,''
  Phys.\ Rev.\ D {\bf 89}, 103008 (2014)
  [arXiv:1403.2049 [hep-th]].
  
  

\bibitem{Rinaldi:2016oqp} 
  M.~Rinaldi,
  ``Mimicking dark matter in Horndeski gravity,''
  Phys.\ Dark Univ.\  {\bf 16}, 14 (2017).
  [arXiv:1608.03839 [gr-qc]].

  
  
  
\bibitem{DeFelice:2010nf}
  A.~De Felice and S.~Tsujikawa,
  ``Generalized Galileon cosmology,''
    Phys.\ Rev.\ D {\bf 84}, 124029 (2011)
[arXiv:1008.4236]].

\bibitem{DeFelice:2011bh}
  A.~De Felice and S.~Tsujikawa,
  ``Conditions for the cosmological viability of the most general
scalar-tensor theories and their applications to extended Galileon dark
energy models,''
  JCAP {\bf 1202}, 007 (2012)
[arXiv:1110.3878]].

\bibitem{DeFelice:2010pv}
  A.~De Felice and S.~Tsujikawa,
  ``Cosmology of a covariant Galileon field,''
   Phys.\ Rev.\ Lett.\  {\bf 105}, 111301 (2010)
[arXiv:1007.2700]]. 

\bibitem{Appleby:2011aa}
  S.~A.~Appleby and E.~V.~Linder,
  ``The Paths of Gravity in Galileon Cosmology,''
   JCAP {\bf 1203}, 043 (2012)
  [arXiv:1112.1981 [astro-ph.CO]].

\bibitem{Babichev:2007dw} 
  E.~Babichev, V.~Mukhanov and A.~Vikman,
  ``k-Essence, superluminal propagation, causality and emergent geometry,''
  JHEP {\bf 0802}, 101 (2008)
  arXiv:0708.0561 [hep-th]].

\bibitem{Deffayet:2010qz} 
  C.~Deffayet, O.~Pujolas, I.~Sawicki and A.~Vikman,
  ``Imperfect Dark Energy from Kinetic Gravity Braiding,''
  JCAP {\bf 1010}, 026 (2010)
  [arXiv:1008.0048 [hep-th]].
  
  
\bibitem{Adams:2006sv} 
  A.~Adams, N.~Arkani-Hamed, S.~Dubovsky, A.~Nicolis and R.~Rattazzi,
  ``Causality, analyticity and an IR obstruction to UV completion,''
  JHEP {\bf 0610}, 014 (2006)
   [hep-th/0602178].
  
\bibitem{Easson:2013bda} 
  D.~A.~Easson, I.~Sawicki and A.~Vikman,
  ``When Matter Matters,''
  JCAP {\bf 1307}, 014 (2013)
   [arXiv:1304.3903 [hep-th]].
  
\bibitem{Ade:2015xua}
  P.~A.~R.~Ade {\it et al.} [Planck Collaboration],
  ``Planck 2015 results. XIII. Cosmological parameters,''
  Astron.\ Astrophys.\  {\bf 594}, A13 (2016)
 [arXiv:1502.01589 [astro-ph.CO]].





\bibitem{Amendola:1993uh}
  L.~Amendola,
  ``Cosmology with nonminimal derivative couplings,''
  Phys.\ Lett.\ B {\bf 301}, 175 (1993)
  [gr-qc/9302010].



\bibitem{Kolyvaris:2011fk}
  T.~Kolyvaris, G.~Koutsoumbas, E.~Papantonopoulos and G.~Siopsis,
  ``Scalar Hair from a Derivative Coupling of a Scalar Field to the Einstein Tensor,''
  Class.\ Quant.\ Grav.\  {\bf 29}, 205011 (2012),
  [arXiv:1111.0263 [gr-qc]].

\bibitem{Rinaldi:2012vy}
  M.~Rinaldi,
  ``Black holes with non-minimal derivative coupling,''
  Phys.\ Rev.\ D {\bf 86}, 084048 (2012)
  [arXiv:1208.0103 [gr-qc]].


\bibitem{Kolyvaris:2013zfa}
  T.~Kolyvaris, G.~Koutsoumbas, E.~Papantonopoulos and G.~Siopsis,
  ``Phase Transition to a Hairy Black Hole in Asymptotically Flat Spacetime,''
  JHEP {\bf 1311}, 133 (2013),
  [arXiv:1308.5280 [hep-th]].


\bibitem{Koutsoumbas:2015ekk}
  G.~Koutsoumbas, K.~Ntrekis, E.~Papantonopoulos and M.~Tsoukalas,
  ``Gravitational Collapse of a Homogeneous Scalar Field Coupled Kinematically to 
Einstein 
Tensor,''
  Phys.\ Rev.\ D {\bf 95}, no. 4, 044009 (2017)
  [arXiv:1512.05934 [gr-qc]].

\bibitem{Sushkov:2009hk}
  S.~V.~Sushkov,
  ``Exact cosmological solutions with nonminimal derivative coupling,''
  Phys.\ Rev.\  D {\bf 80}, 103505 (2009)
  [arXiv:0910.0980 [gr-qc]].



\bibitem{Gao:2010vr}
  C.~Gao,
  ``When scalar field is kinetically coupled to the Einstein tensor,''
  JCAP {\bf 1006}, 023 (2010)
  [arXiv:1002.4035 [gr-qc]].

\bibitem{Granda:2009fh}
  L.~N.~Granda,
  ``Non-minimal Kinetic coupling to gravity and accelerated expansion,''
  JCAP {\bf 1007}, 006 (2010)
  [arXiv:0911.3702 [hep-th]].

\bibitem{Saridakis:2010mf}
  E.~N.~Saridakis and S.~V.~Sushkov,
  ``Quintessence and phantom cosmology with non-minimal derivative coupling,''
  Phys.\ Rev.\  D {\bf 81}, 083510 (2010)
  [arXiv:1002.3478 [gr-qc]].

\bibitem{Germani:2010gm}
  C.~Germani, A.~Kehagias,
  ``New Model of Inflation with Non-minimal Derivative Coupling of Standard Model
Higgs Boson to Gravity,''
  Phys.\ Rev.\ Lett.\  {\bf 105}, 011302 (2010)
  [arXiv:1003.2635 [hep-ph]].
  



  
  

\bibitem{Dent:2013awa} 
  J.~B.~Dent, S.~Dutta, E.~N.~Saridakis and J.~Q.~Xia,
  ``Cosmology with non-minimal derivative couplings:perturbation analysis and 
observational constraints,''
  JCAP {\bf 1311}, 058 (2013)
  [arXiv:1309.4746 [astro-ph.CO]].

  
  
\bibitem{Dalianis:2016wpu}
  I.~Dalianis, G.~Koutsoumbas, K.~Ntrekis and E.~Papantonopoulos,
  ``Reheating predictions in Gravity Theories with Derivative Coupling,''
  JCAP {\bf 1702}, no. 02, 027 (2017)
  [arXiv:1608.04543 [gr-qc]].

\bibitem{Koutsoumbas:2013boa}
  G.~Koutsoumbas, K.~Ntrekis and E.~Papantonopoulos,
  ``Gravitational Particle Production in Gravity Theories with Non-minimal Derivative 
Couplings,''
  JCAP {\bf 1308}, 027 (2013),
  [arXiv:1305.5741 [gr-qc]].






\bibitem{Kuang:2016edj}
  X.~M.~Kuang and E.~Papantonopoulos,
  ``Building a Holographic Superconductor with a Scalar Field Coupled Kinematically to 
Einstein 
Tensor,''
  JHEP {\bf 1608}, 161 (2016),
  [arXiv:1607.04928 [hep-th]].


\bibitem{Bogdanos:2009uj} 
  C.~Bogdanos and E.~N.~Saridakis,
  ``Perturbative instabilities in Horava gravity,''
  Class.\ Quant.\ Grav.\  {\bf 27}, 075005 (2010)
  [arXiv:0907.1636 [hep-th]].
  
\bibitem{DeFelice:2012mx} 
  A.~De Felice, A.~E.~Gumrukcuoglu and S.~Mukohyama,
  ``Massive gravity: nonlinear instability of the homogeneous and isotropic universe,''
  Phys.\ Rev.\ Lett.\  {\bf 109}, 171101 (2012)
  [arXiv:1206.2080 [hep-th]].
  
\bibitem{Afshordi:2006ad} 
  N.~Afshordi, D.~J.~H.~Chung and G.~Geshnizjani,
  ``Cuscuton: A Causal Field Theory with an Infinite Speed of Sound,''
  Phys.\ Rev.\ D {\bf 75}, 083513 (2007)
    [hep-th/0609150].
   
  
\bibitem{Bento:2003we}
  M.~C.~Bento, O.~Bertolami and A.~A.~Sen,
  ``WMAP Constraints on the Generalized Chaplygin Gas Model,''
  Phys.\ Lett.\  B {\bf 575}, 172 (2003)
  [arXiv:astro-ph/0303538].

  
\bibitem{Kamenshchik:2001cp} 
  A.~Y.~Kamenshchik, U.~Moschella and V.~Pasquier,
  ``An Alternative to quintessence,''
  Phys.\ Lett.\ B {\bf 511}, 265 (2001)
  [gr-qc/0103004].

\bibitem{Gorini:2005nw} 
  V.~Gorini, A.~Kamenshchik, U.~Moschella, V.~Pasquier and A.~Starobinsky,
  Phys.\ Rev.\ D {\bf 72}, 103518 (2005)
  [astro-ph/0504576].

    
\bibitem{Granda:2011zy}
  L.~N.~Granda, E.~Torrente-Lujan and J.~J.~Fernandez-Melgarejo,
  ``Non-minimal kinetic coupling and Chaplygin gas cosmology,''
  Eur.\ Phys.\ J.\ C {\bf 71}, 1704 (2011)
  [arXiv:1106.5482 [hep-th]].

  
\bibitem{ArmendarizPicon:2000dh} 
  C.~Armendariz-Picon, V.~F.~Mukhanov and P.~J.~Steinhardt,
  ``A Dynamical solution to the problem of a small cosmological constant and late time 
cosmic acceleration,''
  Phys.\ Rev.\ Lett.\  {\bf 85}, 4438 (2000)
  [astro-ph/0004134].
  
\bibitem{Pujolas:2011he} 
  O.~Pujolas, I.~Sawicki and A.~Vikman,
  ``The Imperfect Fluid behind Kinetic Gravity Braiding,''
  JHEP {\bf 1111}, 156 (2011)
    [arXiv:1103.5360 [hep-th]].
  
  
\bibitem{Harko:2016xip}
  T.~Harko, F.~S.~N.~Lobo, E.~N.~Saridakis and M.~Tsoukalas,
  ``Cosmological models in modified gravity theories with extended nonminimal derivative 
couplings,''
  Phys.\ Rev.\ D {\bf 95}, no. 4, 044019 (2017)
  [arXiv:1609.01503 [gr-qc]].

\bibitem{Suzuki:2011hu}
  N.~Suzuki {\it et al.},
  ``The Hubble Space Telescope Cluster Supernova Survey: V. Improving the Dark Energy
Constraints Above z$>$1 and Building an Early-Type-Hosted Supernova Sample,''
  Astrophys.\ J.\  {\bf 746}, 85 (2012)
 [arXiv:1105.3470 [astro-ph.CO]].
  
  
  \bibitem{Akaike1974} 
H. Akaike, ``A new look at the statistical model identification'', IEEE Transactions 
on 
Automatic Control, 
\textbf{19},  716 (1974).
  

\bibitem{Nunes:2016drj} 
  R.~C.~Nunes, S.~Pan, E.~N.~Saridakis and E.~M.~C.~Abreu,
  ``New observational constraints on $f(R)$ gravity from cosmic chronometers,''
  JCAP {\bf 1701}, no. 01, 005 (2017)
  [arXiv:1610.07518 [astro-ph.CO]].


\bibitem{delaCruz-Dombriz:2016bqh} 
  A.~de la Cruz-Dombriz, P.~K.~S.~Dunsby, O.~Luongo and L.~Reverberi,
  ``Model-independent limits and constraints on extended theories of gravity from cosmic 
reconstruct
ion techniques,''
  JCAP {\bf 1612}, no. 12, 042 (2016)
 [arXiv:1608.03746 [gr-qc]].
  
  
  

 \bibitem{Germani:2010ux} 
  C.~Germani and A.~Kehagias,
  ``Cosmological Perturbations in the New Higgs Inflation,''
  JCAP {\bf 1005}, 019 (2010)
 [arXiv:1003.4285 [astro-ph.CO]].

\bibitem{TheLIGOScientific:2017qsa} 
  B.~P.~Abbott {\it et al.} [LIGO Scientific and Virgo Collaborations],
  ``GW170817: Observation of Gravitational Waves from a Binary Neutron Star Inspiral,''
  Phys.\ Rev.\ Lett.\  {\bf 119}, no. 16, 161101 (2017)
  [arXiv:1710.05832 [gr-qc]].

  
\bibitem{Monitor:2017mdv} 
  B.~P.~Abbott {\it et al.} [LIGO Scientific and Virgo and Fermi-GBM and INTEGRAL 
Collaborations],
  ``Gravitational Waves and Gamma-Rays from a Binary Neutron Star Merger: GW170817 and 
GRB 170817A,''
  Astrophys.\ J.\  {\bf 848}, no. 2, L13 (2017)
  [arXiv:1710.05834 [astro-ph.HE]].


\bibitem{Ezquiaga:2017ekz} 
  J.~M.~Ezquiaga and M.~Zumalacárregui,
  ``Dark Energy after GW170817,''
    Phys.\ Rev.\ Lett.\  {\bf 119}, no. 25, 251304 (2017)
  [arXiv:1710.05901 [astro-ph.CO]].

\bibitem{Baker:2017hug} 
  T.~Baker, E.~Bellini, P.~G.~Ferreira, M.~Lagos, J.~Noller and I.~Sawicki,
  ``Strong constraints on cosmological gravity from GW170817 and GRB 170817A,''
  Phys.\ Rev.\ Lett.\  {\bf 119}, no. 25, 251301 (2017)
  [arXiv:1710.06394 [astro-ph.CO]].

  
  
  

  

  


  
  
  
  
  
  
  

  

  

\end{thebibliography}
\end{document}